# Attosecond inner-shell lasing at Ångström wavelengths

Thomas M. Linker[1,2*], Aliaksei Halavanau[3], Thomas Kroll[4], Andrei Benediktovitch[5], Yu Zhang[1], Yurina Michine[6], Stasis Chuchurka[5], Zain Abhari[2], Daniele Ronchetti[5,7,8], Thomas Fransson[1,9], Clemens Weninger[10,11], Franklin D. Fuller[10], Andy Aquila[10], Roberto Alonso-Mori[10], Sébastien Boutet[10], Marc W. Guetg[3], Agostino Marinelli[1,3], Alberto A. Lutman[3], Makina Yabashi[12,13], Ichiro Inoue[5,12], Taito Osaka[12], Jumpei Yamada[12], Yuichi Inubushi[12], Gota Yamaguchi[12], Toru Hara[12], Ganguli Babu[14], Devashish Salpekar[14], Farheen N. Sayed[14], Pulickel M. Ajayan[14], Jan Kern[15], Junko Yano[15], Vittal K. Yachandra[15], Matthias F. Kling[1,8,16], Claudio Pellegrini[3], Hitoki Yoneda[6], Nina Rohringer[5,7], Uwe Bergmann[2**]

1. Stanford PULSE Institute. SLAC National Accelerator Laboratory, Menlo Park, California 94025, US

2. Department of Physics, University of Wisconsin-Madison, Madison, WI, 53706, USA

3. Accelerator Research Division, SLAC National Accelerator Laboratory, Menlo Park, CA 94025

4. Stanford Synchrotron Radiation Lightsource, SLAC National Accelerator Laboratory, Menlo Park, CA, 94025, USA5.

5. Center for Free-Electron Laser Science CFEL, Deutsches Elektronen-Synchrotron DESY, Notkestr. 85, 22607 Hamburg, Germany

6. Institute for Laser Science, The University of Electro-Communications, Chofu, Tokyo 182-8585, Japan

7. Department of Physics, Universität Hamburg, Hamburg 22761, Germany

8. Max Planck School of Photonics, Friedrich-Schiller University of Jena, Jena 07745, Germany

9. Department of Theoretical Chemistry and Biology, KTH Royal Institute of Technology, Stockholm, Sweden

10. Linac Coherent Light Source, SLAC National Accelerator Laboratory, Menlo Park, California 94025, USA

11. MAX IV Laboratory, Lund University, Lund 224 84, Sweden

12. RIKEN SPring-8 Center, Sayo-cho, Sayo-gun, Hyogo 679-5148, Japan

13. Japan Synchrotron Radiation Research Institute, Sayo-cho, Sayo-gun, Hyogo 679-5198, Japan

14. Department of Materials Science and Nano Engineering, Rice University, Houston, TX 7700512, USA

15. Molecular Biophysics and Integrated Bioimaging Division, Lawrence Berkeley National Laboratory, Berkeley, California 94720, USA

16. Department of Applied Physics, Stanford University. Stanford, California 94305, USA.

**Corresponding Authors**

\* tlinker@slac.stanford.edu

\*\*ubergmann@wisc.edu

**Since the invention of the laser, nonlinear effects such as filamentation[1], Rabi cycling[2,3] and collective emission[4] have been explored in the optical regime leading to a wide range of scientific and industrial applications[5–8]. X-ray free electron lasers (XFELs) have led to the extension of many optical techniques to X-rays for their advantages of Ångström scale spatial resolution and elemental specificity[9]. One such example is XFEL-driven population inversion of 1s core hole states resulting in inner-shell K$\alpha_1$ ($2p_{3/2} \rightarrow 1s_{1/2}$) X-ray lasing in elements ranging from neon to copper, which has been utilized for nonlinear spectroscopy and development of next generation X-ray laser sources[10–16]. Here we show that strong lasing effects, similar to those observed in the optical regime, can occur at 1.5 – 2.1 Å wavelengths during high intensity (>$10^{19}$ W/cm$^2$) XFEL-driven inner-shell lasing and superfluorescence of copper and manganese. Depending on the temporal substructure of the XFEL pump pulses, the resulting inner-shell X-ray laser pulses (containing ~$10^6$-$10^8$ photons) can exhibit strong spatial inhomogeneities as well as spectral splitting, inhomogeneities, and broadening. Through 3D Maxwell Bloch theory[17] we show that the observed spatial inhomogeneities result from X-ray filamentation, and that the spectral**

**splitting and broadening is driven by Rabi cycling with sub-femtosecond periods. Our simulations indicate that these X-ray pulses can have pulse lengths of less than 100 attoseconds and coherence properties that open the door for quantum X-ray optics applications.**

Strong lasing effects such as filamentation and Rabi cycling have allowed us to develop optical techniques for a wide range of uses from fundamental studies of electrons, atoms, molecules, and materials, to cutting-edge scientific and industrial laser applications[5–8,18–20]. There has been a growing experimental effort to extend optical nonlinear and strong-field techniques to the X-ray regime with intense ultrashort pulses provided by X-ray free electron lasers (XFELs). Such work includes the generation of multiple core holes[21–28], X-ray-optical wave mixing[29], and X-ray second harmonic generation[30]. There has even been recent work to extend XFEL experiments to the attosecond regime through development of novel bunch compression schemes and soliton-like amplification methods, leading to next generation attosecond X-ray science[31–39]. Despite these great successes, investigation of higher-order nonlinear processes like filamentation and Rabi cycling have been lacking, especially in the hard X-ray regime i.e. at Ångström wavelengths. Extending nonlinear techniques to hard X-rays is essential for samples requiring the large penetration depth, spatial resolution and high fluorescence yield offered by hard X-ray excitations[40]. In the optical regime, the physical origin of filamentation and its associated spectral broadening originates from 3rd order sample nonlinearities predominantly arising from the Kerr effect and plasma generation that modify the refractive index[7]. In the case of hard X-rays, the index of refraction is very close to unity, and filamentation is traditionally not expected to occur. Rabi cycling[2]—the periodic modulation of populations in two-level systems within a time-varying field—has been observed in the extreme ultraviolet (XUV) and soft X-ray range, but not been reported at Ångström wavelengths.

Here, we report the observation and description of laser filamentation and Rabi cycling in superflourescent inner-shell lasing at the 5.9 keV (2.2Å) Mn K$\alpha_1$ and 8.05 keV (1.5Å) Cu K$\alpha_1$ lines. Angular profiles of the stimulated emission analogous to those seen in laser filamentation are observed and determined to be formed from a balance of spontaneous emission, population inversion and self-amplification with diffraction. Direct observation of Rabi frequencies greater than 5 eV are extracted through the self-induced Autler-Townes[44,45] splitting of the stimulated emission spectra, indicating X-ray-induced coherent attosecond population dynamics. Our simulations show that these strong lasing phenomena can lead to the generation of isolated attosecond X-ray pulses, even if the XFEL self-amplified spontaneous emission (SASE) pump pulse is much longer.

**Stimulated X-ray emission**

The experiments were performed at the nanofocus instrument in Experimental Hutch 5 (EH5) at beamline 3 at the SPring-8 Ångström Compact free electron Laser (SACLA) in Hyogo, Japan and at the Coherent X-ray Imaging (CXI) instrument at the Linac Coherent Light Source (LCLS) at SLAC National Accelerator Laboratory. Both instruments employed highly focused (100-150 nm diameter), intense (~$10^{19}$-$10^{20}$ W/cm$^2$) SASE XFEL pulses respectively tuned above the K edges of Mn (6.54 keV) and Cu (8.98 keV). Spectral and angular analysis of the emission signal was performed using a flat Si analyzer crystal dispersing the emission signal onto a two-dimensional (2D) charge-coupled device (CCD) detector with one 'spectral' axis and one

'spatial/angular' axis depending on how the diffraction plane of the Si analyzer was aligned to the beam (vertically at LCLS, horizontally at SACLA). The schematics of this geometry, which has been used in previous experiments[12–15], is shown in Fig. 1a. The resulting 2D spectral-angular profiles (2D profiles) illustrated in Fig. 1a show the spectral (dispersive) direction indicating the photon energy of the emission signal vertically, and the spatial (non-dispersive) direction, indicating one angular ($\phi$) direction of the emission signal for each wavelength relative to the forward direction, horizontally. In this setup, a monochromatic emission signal with large angular divergence (in both directions) would only show a horizontal line in the 2D profile, as the Bragg analyzer rejects the angle/wavelength mismatch in the dispersive direction. A broadband emission signal with large angular divergence would give a wide signal in the 2D profile both vertically and horizontally. As will be seen in the single-shot 2D profiles discussed throughout this paper, strong lasing effects can cause spatial and/or spectral inhomogeneities and broadening. A more detailed description of the experimental setup and parameters is provided in the Supplementary Information.

To model stimulated X-ray emission we solve the 3D Maxwell Bloch equations within a continuous variable scheme for the atomic variables as described previously.[17] The stimulated emission process for the K$\alpha_1$ line is diagrammed in Fig. 1b and 1c, where an incident SASE XFEL pulse ionizes the Mn/Cu atoms by simultaneously removing many 1s electrons creating an excited state population inversion. A spontaneously emitted K$\alpha_1$ photon resulting from the relaxation of a 2p$_{3/2}$ electron into the 1s$_{1/2}$ core hole can stimulate the emission of a second K$\alpha_1$ photon along the direction of core hole excited states. This results in a cascade of amplified spontaneous emission of K$\alpha_1$ photons as illustrated in Fig. 1b. As the optical path length increases, the influence of the radiation field on the phases of the atomic transition dipoles increases. This causes the transition dipole moments to synchronize leading to faster and faster emission, shortening the stimulated emission pulses and enhancing their peak power. As in the optical regime[46], this can eventually lead to Rabi cycling and generation of pulses much shorter than the excited state (core-hole) life time. We model the $K\alpha_1$ transition in Mn/Cu with a 6-level density matrix consisting of 2 upper levels $|u\rangle$ corresponding to two possible $1s_{1/2}$ core hole excited states with differing magnetic quantum ($m = \pm 1/2$) and 4 lower levels $|l\rangle$ corresponding to 4 possible $2p_{3/2}$ core hole final states with differing magnetic quantum numbers ($m = \pm 3/2, \pm 1/2$). The initial population of the $|u\rangle$ levels are generated via non-resonant photoionization from the initial neutral ground state $|g\rangle$, as illustrated in Fig. 1c. All relevant ionization, fluorescence, and Auger-Meitner decay channels are then considered for the upper and lower levels (see Supplementary Information and reference[17] for further details).

**Filamentation**

We first focus on stimulated emission signals primarily displaying filamentation qualities without strong spectral broadening. This describes the majority of the data taken at LCLS where the peak intensity in the SASE pump pulses was lower than at SACLA. Figures 2a,b show two examples of 2D profiles for stimulated emission from a NaMnO$_4$ solution measured at LCLS. While in Fig. 2a mainly a single spot is seen along $\phi$, two hotspots are apparent in Fig. 2b. In neither case the dispersive direction shows significant spectral broadening with most of the signal occurring at the $K\alpha_1$ photon energy. Similar 2D profiles, some with even more than two hotspots, where also observed in MnCl$_2$ solutions and Mn foils at LCLS (see Supplementary Fig. S3). To understand how such hotspots can form during stimulated emission, we first simulate the spectra generated by a single weak SASE spike at solution density. The calculated real-space *xy* profile of the stimulated emission at the end of propagation through the medium (*z* axis is the propagation axis) is plotted in Fig. 2c. The *xy* profile displays two hotspots along the *y* axis indicative of a filamentation process. To simulate the 2D profiles of our measured signal, we compute the intensity of the far-field stimulated emission and propagate it through the response function of the Bragg analyzer (see Supplementary Figs. S4 and S5). Figure 2d shows the calculated 2D profile of the stimulated emission, where the $\phi$ axis is aligned with the *x* axis of the *xy* profile shown in Fig. 2c. Only one spot is apparent. Figure 2e shows the calculated 2D profile of the stimulated emission, where the $\phi$ axis is aligned with the *y* axis of the *xy* profile shown in Fig. 2c, and two spots appear. These calculated 2D profiles shown in Fig. 2d,e are representative of the two measured 2D profiles shown in Figs. 2a,b. This shows that filamentation is only detected in the 2D profile, if the hot spots are aligned with the non-dispersive $\phi$ axis. Filamentation hot spots can occur of course in any direction resulting in a variety of inhomogeneities in 2D profiles, but as there is no significant spectral broadening, these inhomogeneities occur mainly along the $\phi$ axis. As the amplified spontaneous emission process starts from noise, we do not expect simulated 2D profiles to perfectly match any observed 2D profile, but rather to capture the trends.

Figure 2f shows the temporal profile of a simulated Gaussian pump pulse (red) that mimics a single SASE spike (pulse length ~235 attoseconds FWHM, intensity ~$10^{19}$ W/cm$^2$), and the temporal profile of Mn K$\alpha_1$ stimulated emission (blue) spatially integrated over the transverse axes at the end of the pulse propagation. Our simulation indicates the stimulated emission pulse generated from the SASE spike under conditions leading to filamentation has ~600 attoseconds

FWHM pulse length. Snapshots of the filamentation dynamics that lead to the formation of the hotspots are plotted in Fig. 2g, showing several *xy* profiles of the stimulated emission, as it propagates through the sample. At the beginning of the propagation the signal is dominated by spontaneous emission resulting in a wide and random distribution of the emitted radiation in space. As stimulated emission begins to dominate it focuses toward the center of the pump power, experiencing a gain guiding effect that creates a high buildup of photons within a bounded region that can be much smaller than the pump focus. In this bounded region multiple hotspots (filaments) can form as diffraction competes with the field amplification. Upon exhaustion of the pump, the dynamics become diffraction dominated resulting in a nearly free propagation regime (see also Supplementary Video S1).

In both data and simulations, we find the distribution of angular hotspots for a given pump pulse to be highly variable from shot to shot, as these hotspots are determined by quantum fluctuations that seed the spontaneous emission process (see Supplementary Figs. S6-S9 for more discussion). In the vast majority of measured shots, we detect only a single amplified mode (see Supplementary Fig. S6), but, as shown in Fig. 2d, we are only sensitive to multiple hotspots if they occur along the non-dispersive direction ($y \rightarrow \phi$). Our 3D Maxwell Bloch simulations using the full ~30 fs SASE pump pulses (with multiple spikes) employed in the experiments at LCLS, show that the ground state is completely depopulated on the rise of the pump pulse (for more discussion see Supplementary Figs. S10). This important finding explains why using a single spike to represent the SASE pump pulse in our simulations can well reproduce the 2D profiles measured at LCLS.

**Rabi cycling and spectral broadening**

We next examine the transition into the stronger lasing regime, where spectral broadening occurs. We focus our discussion on spectral broadening and self-induced Autler-Townes splitting noting that we have also observed the formation of Mollow triplets (see discussion in Supplementary Information and Figs. S11 and S12). Figure 3a shows the spectral width versus number of photons for 29,000 single-shot spectra from $NaMnO_4$ taken at LCLS . A constant spectral width over several orders of magnitude of gain, indicative of the buildup transform limited pulses (see ref[12]), is followed by strong spectral broadening beginning at $\sim 10^6$ photons/pulse in the saturation regime. Deep in saturation, the main signal is redshifted exhibiting two peaks, as shown in Fig. 3b. (It is important to note that the $K\alpha_2$ emission signal is strongly suppressed in stimulated emission[12], and would appear at 5887 eV for $NaMnO_4$.) The measured 2D profile (top) can be simulated using a $NaMnO_4$ solution pumped by a high intensity single SASE spike displaying the same peak splitting and shifts (bottom). The corresponding pulse length is 320 attoseconds FHWM (see Supplementary Fig. S17). Shifts/broadening due to the electronic structure of the medium can be ruled out for the origin of the shift and splitting (see Supplementary Figs. S13-S17 and ref[12]). Furthermore, we only see such pronounced splitting and red shift when entering the strong saturation regime (see Supplementary Figs. S13-S17). Our simulations show that this spectral feature is a result of the self-induced Autler-Townes splitting, consistent with previous theory work[45]. The splitting of ~3.5 eV observed in the spectra measured at LCLS corresponds to the Rabi frequency. At SACLA, where the pump pulse intensity was higher, we observed larger Autler-Townes splittings of ~5 eV for a 7 μm copper foil and the corresponding simulation yields a 230 attosecond FWHM pulse length (see Supplementary Fig. S19). Such splitting only occurs when the population is inverted on a time scale much shorter than the core-hole lifetime,[45] which further indicates that these strong lasing shots are driven by a single strong spike from the SASE pump pulse.

We now show that the observed asymmetry of the Autler-Townes splitting requires both forward and transverse radiation propagating in a highly inverted gain medium. Fig. 3c shows the calculated field intensity and phase dynamics for stimulated emission in the high-gain regime (high density and high pump intensity) at the focus of the pump pulse as well as the fully spatially integrated spectral intensity as function of the propagation axis. The top row shows the dynamics when transverse propagation of the fields is not included (1D dynamics) and the bottom row when it is included (3D dynamics). Both cases exhibit intensity ringing with stronger effects in the 1D case. Similar to the optical regime[46], the ringing is much weaker than the main emission pulse, and the main emission pulse length is much shorter than the excited state (core-hole) lifetime. For the 1D dynamics we see the expected $\pi$ phase shift in the intensity ringing during Rabi cycling and a corresponding symmetric spectral profile that is not consistent with our measured 2D profiles. For the 3D case the phase dynamics are much more complex upon entering the Rabi cycling regime. This results in intensity ringing with a time dependent phase, generating a non-symmetric spectral profile, consistent with our measured 2D profiles. We find that nevertheless the phase is nearly constant within the FWHM of the stimulated emission pulse (see Supplementary Fig. S19). Driven deep into saturation, the Autler-Townes spitting becomes washed out and the spectrum becomes broad and inhomogeneous when 3D dynamics are considered. For such cases our simulations indicate the pulse lengths are of order 100 attoseconds. This is the case for a large fraction of the lasing shots obtained at SACLA, where we found that 78% of all stimulated emission shots from a 20 μm thick Cu metal foil displayed broad inhomogeneous spectra (see Supplementary Fig. S20).

Figure 4a shows an example of a shot deep in saturation taken at SACLA for a MnO loaded carbon film. The 2D profile exhibits spatial and spectral inhomogeneities and large spectral broadening. For comparison, a simulation at similar conditions is shown in Fig. 4b. Figure 4c shows the temporal profile of the SASE pump pulse (blue dashed) used for the simulation and the temporal response of the stimulated emission signal (red) resulting in a pulse length of 120 attoseconds FWHM. Here, a single spike of the SASE pump pulse completely depopulates the ground state and drives the emission deep into saturation. The Autler-Townes splitting is washed out resulting in a broad and inhomogeneous spectrum. We note the simulation is slightly more spatially inhomogeneous than the experiment. We attribute this to the simulation being performed at the sample average density, whereas the film density was not uniform (see Supplemental Information). Figures 4d and 4g show further examples of 2D profiles of single shots deep in saturation for a MnO loaded film and a Cu 20$\mu$m foil. In addition to the spatial and spectral inhomogeneities and large spectral broadening, spectral fringes with regular spacing start to appear in Fig. 4d, and they can be seen more prominently in Fig 4g. We have previously observed and described 2D profiles exhibiting well-defined spectral fringes for Mn stimulated emission[15]. Such well-defined fringes result from the interference of temporally coherent stimulated emission pulse pairs generated from two separate SASE spikes within the pump pulse. The fringe spacing $\varDelta E$ is related to the pulse spacing $\varDelta t$ through the Fourier analysis in the Bragg spectrometer with $\varDelta t \varDelta E = h = 4.136$ fs-eV, where $h$ is the Planck constant. The fringe contrast and envelop depends on the pulse length, coherence, and relative strengths of the two pulses. Figures 4e and 4h show simulations of 2D profiles exhibiting fringes comparable to the measured spectra (Figs. 4d and 4g). Figures 4f and 4i, show the corresponding temporal profiles of SASE pump pulses (blue dashed) used for the simulations, and the temporal profiles of the stimulated emission signals (red). The pulse lengths of the first pulses in the simulated emission signals are 100 and 90 attoseconds FWHM respectively, as shown in Fig. 4f and 4i. Figure 4f shows that the second emission pulse is 20 times weaker than the first one, and yet weak fringe contrast can still be seen in the corresponding 2D profile in Fig. 4e. When the second emission pulse has similar intensity and pulse length to the first pulse (Fig. 4i), the corresponding 2D profile is broad and displays fringes with deep contrast (Fig. 4h). In this case the stimulated emission temporal profile has high intensity displaying pulses with 100 attoseconds FWHM length.

We also observed spectrally broad 2D profiles showing deep fringes with spacings up to 8 eV corresponding to pulse separations of ~500 attoseconds (see Supplementary Fig. S21). As shown in Fig 4i, multiple deep fringes are indicative of two stimulated emission pulses with similar intensity and length. This is consistent with these pulses having a nearly constant phase (see Supplementary Fig. S19). In this case, multiple deep interference fringes can only occur if the length of each pulse in the pair is much shorter than the corresponding temporal spacing, i.e. much shorter than 500 attoseconds for an 8 eV fringe spacing.

**Conclusions**

We have demonstrated how stimulated emission of hard X-rays can behave attune to an optical laser in the strong lasing regime. Our simulations based on 3D Maxwell-Bloch theory[17] show that spatial filamentation is driven through gain guiding effects in the transition from spontaneous to stimulated emission. Spectral broadening is driven by the self-compression of the stimulated emission pulse from Rabi cycling with sub-fs periods. Such pulse compression is a key feature of the collective emission during superfluorescence, where shortening of pulses is accompanied by an increase in their peak power [46]. Rabi cycling is the fundamental building block for most studies of coherent control of quantum systems. In many observed 2D profiles, the extracted Rabi frequencies are greater than 5 eV (<0.86 fs) indicating our ability to drive coherent attosecond inner-shell population dynamics. Controlling these dynamics will be essential for novel source development and the next generation X-ray spectroscopy and quantum optics applications. The recent development of single spike SASE XFEL pulses[37,39] can help to better control these dynamics, and heavier gain mediums with shorter lifetimes (e.g. Lα emission of W or Hf at ~8 keV) can potentially generate even shorter pulses well below 100 attoseconds. We are currently working on schemes to spatially and/or spectrally separate the stimulated emission from the pump pulse. As the angular divergence of the stimulated emission pulses is determined by the medium density and overall gain length of the sample its divergence can be several mrad, which is larger than the divergence of a strongly focused pump pulse. This makes it possible to separate the signal spatially from the pump pulse by choosing an interaction point further downstream. Spectral separation can be achieved using a downstream filter with an absorption edge between the photon energies of the pump pulse and the stimulated emission. For a Cu gain medium, a 50 μm thick Cu foil transmits ~10% of the ~8 keV Kα$_1$ stimulated emission pulse while transmitting only ~3 ×10$^{-6}$ of the 9 keV pump pulse.

**Acknowledgements :**

We acknowledge the support from the LCLS and SACLA accelerator group and their technical and engineering staff. Special thanks go to Matthew Seaberg and Matthew Hayes from the CXI instrument. This work is supported by the U.S. Department of Energy (DOE), Office of Science, Basic Energy Sciences (BES) under contract Nos. DE-SC0023270 (UB); DE-SC0009914, DE-AC02-76SF00515(UB), DE-SC-0023585 (AH, UB, CP), and DE-SC-0063 (MFK); Office of Basic Energy Sciences (OBES), Division of Chemical Sciences, Geosciences, and Biosciences (CSGB) of the Department of Energy (DOE) Contract No. DE-AC02-05CH11231 (JYano, VKY); the National Institutes of Health (NIH) Grants No. GM149528 (VKY), No. GM110501 (JYano), No. GM126289 (JK), and the Ruth L. Kirschstein National Research Service Award (F32GM116423, FDF). The experiment at SACLA was performed with the approval of the Japan Synchrotron Radiation Research Institute (proposal Nos. 2017B8066 & 2024A8055) JPSJ KAKENHI is acknowledged for Grant No. 19K20604 (II). SSRL Structural Molecular Biology Program is supported by the DOE Office of Biological and Environmental Research, and by the National Institutes of Health, National Institute of General Medical Sciences (including P41GM103393) (TK). The contents of this publication are solely the responsibility of the authors and do not necessarily represent the official views of NIGMS or NIH (TK). Computer resources for simulations were provided by the National Energy Research Scientific Computing Center (NERSC), a U.S. Department of Energy Office of Science User Facility located at Lawrence Berkeley National Laboratory, operated under Contract No. DEAC02-05CH11231 using NERSC award ERCAP0020725. A. B., S.C. and N.R. acknowledge support from DESY, a member of the Helmholtz Association HGF. S.C. acknowledges the financial support of Grant-No.~HIDSS-0002 DASHH (Data Science in Hamburg-Helmholtz Graduate School for the Structure of Matter). This work is supported by the Cluster of Excellence 'CUI: Advanced Imaging of Matter' of the Deutsche Forschungsgemeinschaft (DFG) - EXC 2056 - project ID 390715994.

**Author Contributions :**

N.R and U.B conceived the original research. T.M.L., T.K, Y.Z., Y.M., Z.A., D.R., I.I., T.O, J.Y.(1), Y.I., G.Y., T.H., V.K.Y., H.Y., and U.B. performed the experiments at the SACLA; T.K, T.F. ,C.W., F.D.F, A.A, R.A-M, S.B, M.W.G, A.A.L., V.K.Y., and U.B. performed the experiments at LCLS; G.B., D.S., F.N.S. and P.M.A. provided the Mn-loaded polymer composites for experiments at SACLA; Y.M. and H.Y. provided the Cu Sulfate pentahydrate and Cu acetate samples for the experiments at SACLA. T.M.L. performed and analyzed the 3D Maxwell Bloch simulations supported by discussions with A.H. and A.B. T.M.L. analyzed the experimental data supported by discussions with A.H., A.B., and U.B.; T.M.L and U.B wrote the first draft of the manuscript; all authors participated in discussions and editing of the manuscript.

Figures :

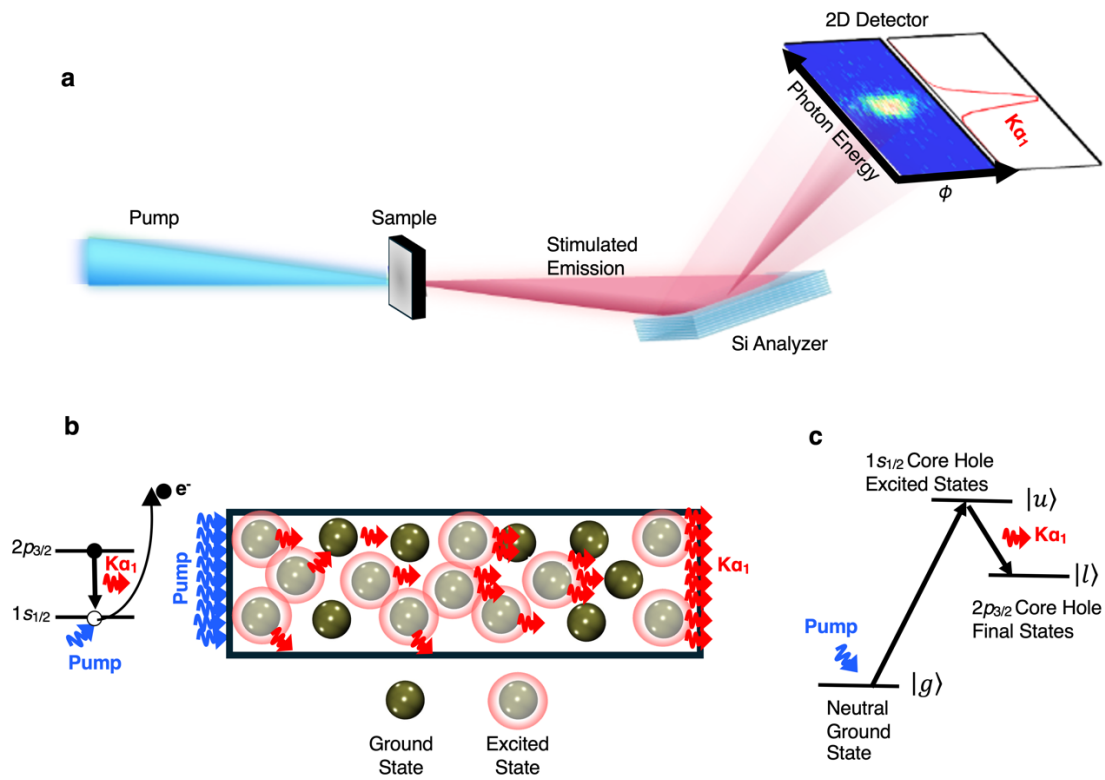

**Figure 1. Experimental setup and concept of stimulated emission.** (a) The highly directional stimulated emission signal is analyzed with a flat Si(220) crystal with the Bragg angle centered at the Mn or Cu Kα₁ followed by a 2D CCD detector. (b) Stimulated emission process is initiated by a SASE pump pulse ejecting many 1s electrons leading to emission of Kα₁ photons when $2p_{3/2}$ electrons fill the $1s_{1/2}$ core hole. During stimulated emission, Kα₁ photons emitted along the forward direction stimulate emission of more Kα₁ photons resulting in exponential gain. (c) State diagram employed in 3D Maxwell-Bloch theory to simulate stimulated emission process. Atoms are excited from an initial ground state $|g\rangle$ to set of upper levels $|u\rangle$ corresponding to the two possible $1s_{1/2}$ core hole states with differing magnetic quantum ($m = \pm 1/2$). The Kα₁ transition is to a set of 4 lower levels $|l\rangle$ corresponding to 4 possible $2p_{3/2}$ core hole final states with differing magnetic quantum numbers ($m = \pm 3/2, \pm 1/2$) .

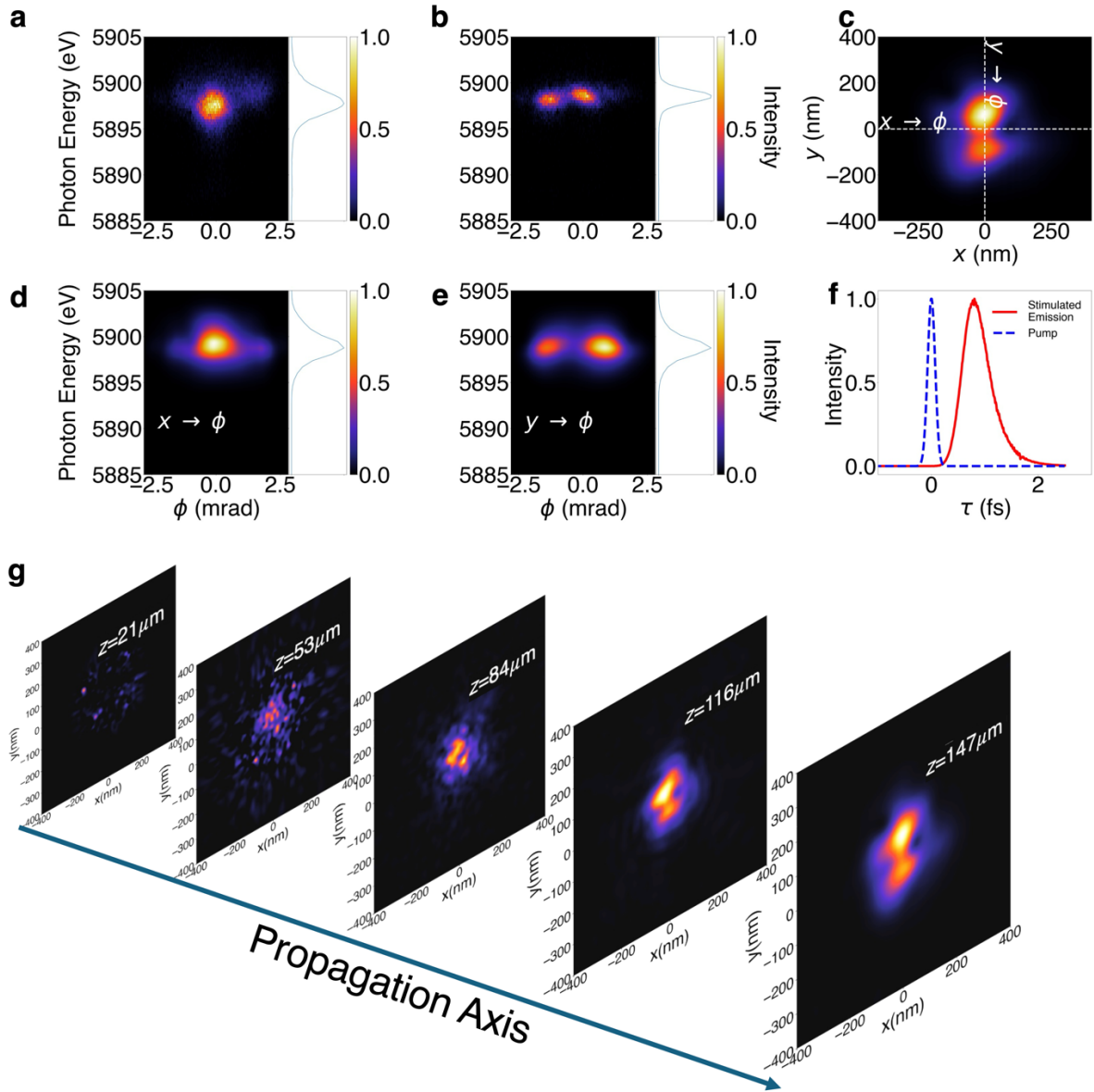

**Figure 2. Filamentation of stimulated Emission**. (a, b) 2D profiles taken at LCLS for NaMnO$_4$ displaying hotspots in their spatial direction ($\phi$). (c) Real space ($xy$) simulation of stimulated emission leaving the medium with 2 hot spots along the $y$ axis. (d, e) Simulation of 2D profiles based on (c) with the spatial direction ($\phi$) corresponding to either the $x$ axis (d) or the $y$ axis (e). The figures show that hot spots are more dominant in the 2D profile when they are aligned with the spatial direction ($\phi$). (f) Temporal profile of pump (dashed) and stimulated emission (red) for simulation shown in (c). (g) Snap shots depicting $xy$ profile of the stimulated emission simulation shown in (c) as it propagates in the gain medium showing the self-focusing and filamentation process. (See Supplementary Information for full video.)

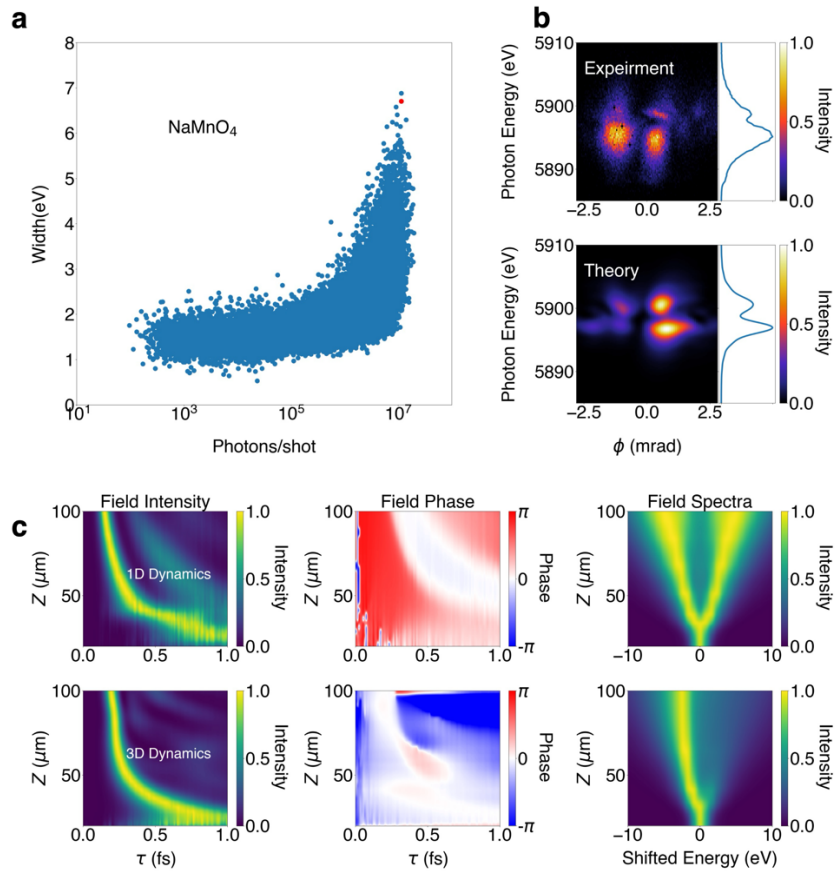

**Figure 3. Transition to strong lasing and Rabi cycling**. (a) Measured spectral width versus number of stimulated emission photons indicating transition to strong lasing regime. (b) Experimental 2D profile for NaMnO$_4$ solution in the strong lasing regime (red point in a) with shoulder peak indicative of Autler-Townes splitting (top) compared to simulation showing same asymmetric splitting (bottom). (c) Phase dynamics of the stimulated emission. Top: 1D simulation showing symmetric splitting with equal intensity in each emission peak. Bottom: 3D simulation showing modulation during the Rabi cycling resulting in asymmetric, red-shifted spectrum. The Rabi ringing is no longer temporally coherent with the parent emission spike indication self-phase modulation of the radiation during 3D propagation. This asymmetry is consistent with the experimental 2D profile (b).

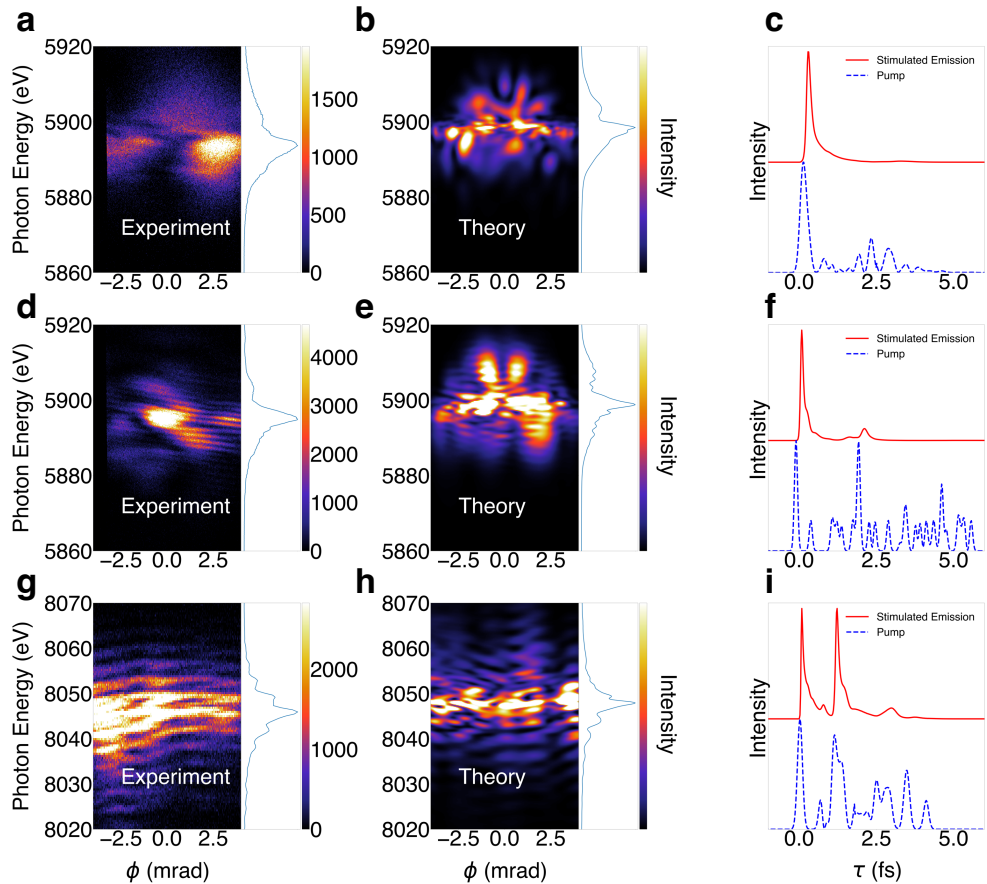

**Figure 4. Broadening through Rabi Cycling at High Intensities**. (a) Experimental 2D profile displaying large spatial and spectral inhomogeneities/broadening for MnO film. (b) Simulation of 2D profile showing similar features using a SASE pump pulse and same average Mn density as in the experiment (a). (c) Temporal profile of the SASE pump pulse (dashed) and stimulated emission signal (red) used in the simulation showing 120 attosecond FWHM pulse length. (d) Experimental 2D profile displaying large spatial and spectral inhomogeneities/broadening and the onset of spectral fringes with ~2eV spacing for MnO film. (e) Simulation of 2D profile showing similar features using a SASE pump pulse and same Mn density as in the experiment (d). (f) Temporal profile of the SASE pump pulse (dashed) and stimulated emission signal (red) used in the simulation (e). The resulting signal shows a strong pulse with 100 attosecond FWHM length, and a much smaller second pulse, delayed by ~2 fs, corresponding the fringe spacing. (g) Experimental 2D profile displaying large spatial and spectral inhomogeneities/broadening and the more pronounced spectral fringes with ~4 eV spacing for Cu foil. (h) Simulation of 2D profile showing similar features using a SASE pump pulse and same Cu density as in the experiment (g). (i) Temporal profile of the SASE pump pulse (dashed) and stimulated emission signal (red) used in the simulation (h). The resulting signal shows two strong pulses with 90 and 100 attosecond FWHM lengths. The pulses are delayed by ~1 fs, corresponding to ~ 4eV fringe spacing.

**Supplemental Information for Attosecond Inner-Shell Lasing at Angstrom Wavelengths**

Thomas M. Linker[1,2*], Aliaksei Halavanau[3], Thomas Kroll[4], Andrei Benediktovitch[5], Yu Zhang[1], Yurina Michine[6], Stasis Chuchurka[5], Zain Abhari[2], Daniele Ronchetti[5,7,8], Thomas Fransson[1,9], Clemens Weninger[10,11], Franklin D. Fuller[10], Andy Aquila[10], Roberto Alonso-Mori[10], Sébastien Boutet[10], Marc W. Guetg[3], Agostino Marinelli[1,3], Alberto A. Lutman[3], Makina Yabashi[12,13], Ichiro Inoue[5,12], Taito Osaka[12], Jumpei Yamada[12], Yuichi Inubushi[12], Gota Yamaguchi[12], Toru Hara[12], Ganguli Babu[14], Devashish Salpekar[14], Farheen N. Sayed[14], Pulickel M. Ajayan[14], Jan Kern[15], Junko Yano[15], Vittal K. Yachandra[15], Matthias F. Kling[1,8,16], Claudio Pellegrini[3], Hitoki Yoneda[6], Nina Rohringer[5,7], Uwe Bergmann[2**]

1. Stanford PULSE Institute. SLAC National Accelerator Laboratory, Menlo Park, California 94025, US

2. Department of Physics, University of Wisconsin-Madison, Madison, WI, 53706, USA

3. Accelerator Research Division, SLAC National Accelerator Laboratory, Menlo Park, CA 94025

4. Stanford Synchrotron Radiation Lightsource, SLAC National Accelerator Laboratory, Menlo Park, CA, 94025, USA5.

5. Center for Free-Electron Laser Science CFEL, Deutsches Elektronen-Synchrotron DESY, Notkestr. 85, 22607 Hamburg, Germany

6. Institute for Laser Science, The University of Electro-Communications, Chofu, Tokyo 182-8585, Japan

7. Department of Physics, Universität Hamburg, Hamburg 22761, Germany

8. Max Planck School of Photonics, Friedrich-Schiller University of Jena, Jena 07745, Germany

9. Department of Theoretical Chemistry and Biology, KTH Royal Institute of Technology, Stockholm, Sweden

10. Linac Coherent Light Source, SLAC National Accelerator Laboratory, Menlo Park, California 94025, USA

11. MAX IV Laboratory, Lund University, Lund 224 84, Sweden

12. RIKEN SPring-8 Center, Sayo-cho, Sayo-gun, Hyogo 679-5148, Japan

13. Japan Synchrotron Radiation Research Institute, Sayo-cho, Sayo-gun, Hyogo 679-5198, Japan

14. Department of Materials Science and Nano Engineering, Rice University, Houston, TX 7700512, USA

15. Molecular Biophysics and Integrated Bioimaging Division, Lawrence Berkeley National Laboratory, Berkeley, California 94720, USA

16. Department of Applied Physics, Stanford University. Stanford, California 94305, USA.

**Corresponding Authors**

* tlinker@slac.stanford.edu

**ubergmann@wisc.edu

*Experimental setup*

Experiments were performed at the nanofocus instrument BL3 at the SACLA XFEL and at the Coherent X-ray Imaging (CXI) hutch at the Linac Coherent Light Source (LCLS) at SLAC National Accelerator Laboratory. The schematics of the experiments have been described previously[1–6]. The pump was focused to a diameter of 100-200 nm at the sample location by Kirkpatrick-Baez (KB) optics, resulting in high peak power of ~$10^{19}$-$10^{20}$ W/cm$^2$ to generate sufficient population inversion required to reach strong saturation of the stimulated emission process. At LCLS, ~30 fs pump pulses with pulse energies ranging from 1-3 mJ and an angular divergence of ~1 mrad horizontally and ~2 mrad vertically were employed. At SACLA, ~6 fs pulses with ~200 µJ and an angular divergence of ~4 mrad horizontally and ~2 mrad vertically were employed. A table of used Samples is provided below :

| Solid Samples | Thickness(um) | density(gcc) |
|---|---:|---:|
| Cu Foil | 7 | 8.96 |
| Cu Foil | 20 | 8.96 |
| CuSO$_4$*5H$_2$O Powder | 77 | 2.286 |
| Cu(CH3COO)$_2$ H$_2$O Powder | 149 | 1.88 |
| Mn Foil | 15 | 7.44 |

| Liquid Jets | Thickness(um) | density(M) |
|---|---:|---:|
| MnCl$_2$ | 200 | 2-5 |
| NaMnO$_4$ | 200 | 2-5 |

| Loaded Carbon Films | Thickness(um) | % metal loading | ~metal density (gcc) |
|---|---|---|---|
| MnSO₄ | 100 | 15 | 0.48 |
| MnO | 100 | 15 | 0.8 |

**Table S1.** Used Samples

The Cu and Mn foil samples were purchased commercially. The 2-5 mol $MnCl_2$ and $NaMnO_4$ solutions were prepared from commercially purchased $MnCl_2$ and $NaMnO_4$ samples from Sigma Aldrich. The solutions at LCLS were delivered into the X-ray beam with a jet using HPLC (high-performance liquid chromatography) and peristaltic pumps with a flow rate of 40 m/s. The MnO, $MnSO_4$ solids were condensed in carbon films at the Ajayan lab at Rice University. The $CuSO_4$ and Cu acetate powders were prepared from commercially produced powders from Sigma Aldrich. The powders had reasonably uniform density, but the condensed films were much more inhomogeneous.

As discussed in previous papers[1–6], each pump pulse destroys the sample, therefore solid samples were raster scanned in an xy scanner and liquid samples were delivered by a jet. In solid samples the spacing between subsequent pulses has to be larger than the crate size from the impact of the pump pulse. In liquid samples, the jet speed has to be larger than the impacted region. We have assured that these conditions were met in all our experiments. Typical crater sizes for a metallic foil are 25-micron diameter (see ref[6]), and we have selected 100 micron distance between pulses for the solid samples to avoid hitting part of the crater caused by the previous shot. Powders were exposed to the beam using kapton tape, with an example shown in figure S1.

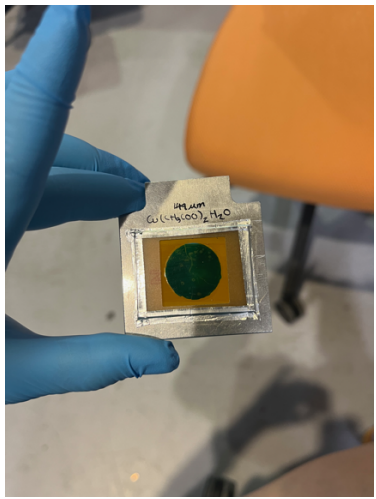

Figure S1. Cu acetate Sample before exposure to the FEL beam.

The SACLA linac was operated at an electron beam energy of 6.03 GeV. XFEL self-amplified spontaneous emission (SASE) pulses are focused through Kirkpatrick-Baez (KB) optics with an estimated 2 mrad vertical divergence and 4 mrad horizonal divergence. The pump pulse energy was ~180 µJ at 6.6 keV/9.1 keV photon energy for Cu/Mn. The beam size was ~200 nm diameter and pulse length of ~6 fs.

A similar experimental setup was used at the LCLS CXI instrument. The incoming X-ray pulses were tuned to 6.6 keV. For the jet samples they were focused onto a liquid jet (200 µm at the beam interaction) with an estimated spot size of ~150 nm diameter and focal depth of ~200 µm. The pulse energies were between 0.5 and 4 mJ measured by an upstream gas detector, with approximately 20% of it reaching the sample after the transport and focusing optics. This corresponds to a maximal XFEL pump intensity at the sample of $1.5\times10^{20}$ W/cm$^2$ for 30 fs pulses. LCLS was operated at an X-ray pulse rate of 30 Hz to ensure jet stability. SACLA was also operated at 30Hz.

To analyze the Kα stimulated emission spectra we employed a Si (220) crystal followed by a 2D MultiPort Charged Coupled Device (MPCCD) detector. We used the CS-PAD 140K[7], which is the same detector that we have used in the previous experiments. The experiments at LCLS were performed in a vertical scattering geometry and those at SACLA in horizontal scattering geometry for Mn and a vertical scattering geometry for Cu. The analyzer had a 1° asymmetric cut to reduce the background from the specular reflection for the Mn experiments.

*Maxwell Bloch Equations.*

To model the stimulated emission we adopt the 3D Maxwell Bloch scheme described in supplemental reference[8], which was first used to investigate $K\alpha_1$ superfluorescence in Cu. Our approach utilizes a continuous variable scheme for the fields and local atomic densities. We have now also created an implementation to Mn. The method for Mn is identical to that described in supplemental reference[8] except changes to the photoionization cross sections, transition dipole moments and X-ray emission energies to that of Mn. For sake of continuity we briefly describe here the method outlined in supplemental reference[8] to highlight which parameters change in this work to reflect that of Mn versus Cu.

We model the $K\alpha_1$ transition in Cu and Mn with a 6 level density matrix with a state diagram illustrated in figure S1. In our analysis throughout the paper, we focus on the emission from the Kα$_1$ transition. Orbital angular momentum splitting of the 2p level ($2p_{3/2}$ and $2p_{1/2}$ levels) leads to the Kα$_2$ line ($2p_{1/2} \rightarrow 1s_{1/2}$ transition). For Mn this transition is ~10 eV lower in energy and ~20eV lower for Cu. The Kα$_2$ line is a factor of 2 weaker in its spontaneous emission spectra. Previous experiments found that in the medium lasing regime, the Kα$_2$ transition is 2-3 orders of magnitude weaker than the Kα$_1$ transition[1] and 1-D Maxwell Bloch simulations have shown a similar intensity ratio in the strong lasing regime.[2] While for the strongest shots we do anticipate the Kα$_2$ to contribute to spectrum, our 3D simulations show that the majority of the spectral features can be attributed to the Kα$_1$ line.

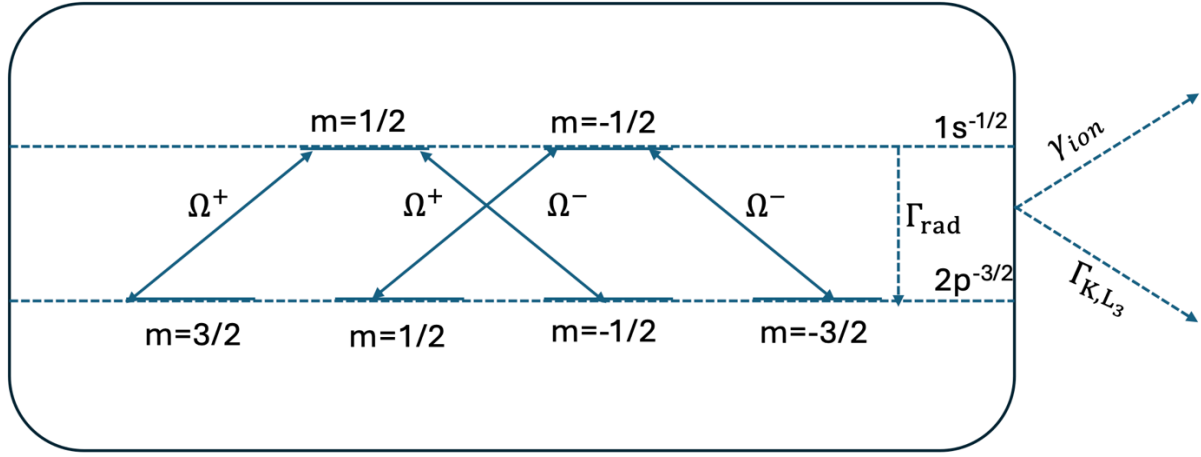

**Figure S2.** State Diagram for Cu/Mn K$\alpha_1$ transition.

We label the 1s$^{-1/2}$ state as the 1s$^{1/2}$ with a core hole and the 2p$^{-3/2}$ as the 2p$^{3/2}$ state containing a core hole. The 1s$^{-1/2}$ can radiatively decay to 4 degenerate 2p$^{-3/2}$ states with spontaneous rate $\Gamma_{rad}$, and are coupled by two radiation modes ($\Omega_{\pm}$), with different polarizations.

The density matrix can be divided into a manifold between upper and lower levels :

$$\{u\} = \left\{1s^{-\frac{1}{2}}_{m=\frac{1}{2}}, 1s^{-\frac{1}{2}}_{m=-\frac{1}{2}}\right\} \tag{1}$$

$$\{l\} = \left\{2p^{-\frac{3}{2}}_{m=\frac{3}{2}}, 2p^{-\frac{3}{2}}_{m=\frac{1}{2}}, 2p^{-\frac{3}{2}}_{m=-\frac{1}{2}}, 2p^{-\frac{3}{2}}_{m=-\frac{3}{2}}\right\} \tag{2}$$

The interaction Hamiltonian between the emission field and the atomic variables is computed in the dipole approximation and depends on matrix elements of the dipole moment operator $\boldsymbol{d_{ul}}$. We consider the 2 initial states within the upper manifold $\{u\}$ and the 4 final states in the lower manifold $\{l\}$ degenerate. We can then write K$\alpha_1$ transition in terms of a single transition strength $d_0$ with

$$\boldsymbol{d_{ul}} \cdot \boldsymbol{e_s} = d_0 T_{ul,s}; \boldsymbol{d_{lu}} \cdot \boldsymbol{e_s^*} = d_0 T_{lu,s} \tag{3}$$

$\boldsymbol{e_s}$ represent the polarization vectors of the field assumed to be perpendicular to the propagation direction (z axis). $T_{ul,s} = T_{lu,s}^*$ store the directional information and are proportional to Clebsch-Gordan coefficients.

$\Gamma_{rad}$ is similarly computed within the dipole approximation for the radiative transition:

$$\Gamma_{rad} = \frac{\omega_0^3 d_0^2}{3\pi\epsilon_0 \hbar c^3} \tag{4}$$

where $\omega_0$ is the fundamental frequency of the K$\alpha_1$ emitted radiation.

All ionic states are created by nonresonant photoionization from the initial neutral ground state. These excited states can decay spontaneously due to Auger-Meitner decay ($\Gamma_K$, $\Gamma_{L3}$), spontaneous emission emitted in other directions other than forward direction, or by further photoionization from the pump and emitted fields. We used $\Gamma_K$=2.24 fs for Cu and 1.69 fs for Mn, and $\Gamma_{L3}$=0.96 fs for Cu and 0.48 fs for Mn. These values were calculated with the Xatom software[9]

In addition to creation of ionic excited states from nonresonant photoionization by the pump field from the initial neutral ground state, we consider further ionization of these ionic states with the pump and emission fields. We explicitly calculate the cross sections for neutral and ionized 1s and 2p levels. We also allow for ionization of the states that do not directly participate in the K$\alpha_1$, but otherwise remove the system from the K$\alpha_1$ excited state manifold, which we account with auxiliary variable (*aux*). Table S1 and S2 tabulate the ionization cross sections Mn for the pump and emission field. The values for Cu are summarized in reference[8].

| Cross section | Value(nm$^{-2}$) | Cross section | Value(nm$^{-2}$) |
|---|---|---|---|
| $\sigma_{1s}^{neutral}$ | 3.57e-6 | $\sigma_{1s}^{ionized}$ | 5.47770e-7 |
| $\sigma_{2p}^{neutral}$ | 1.26e-7 | $\sigma_{2s}^{ionized}$ | 4.03275e-6 |
| $\sigma_{aux}^{neutral}$ | 3.81e-7 | $\sigma_{aux}^{ionized}$ | 3.94303e-6 |

**Table S1.** Ionization cross sections for the pump field for Mn.

| Cross section | Value(nm$^{-2}$) | Cross section | Value(nm$^{-2}$) |
|---|---|---|---|
| $\sigma_{1s}^{neutral}$ | 0 | $\sigma_{1s}^{ionized}$ | 7.52932e-7 |
| $\sigma_{2p}^{neutral}$ | 1.85e-7 | $\sigma_{2s}^{ionized}$ | 6.53510e-7 |
| $\sigma_{aux}^{neutral}$ | 5.07e-7 | $\sigma_{aux}^{ionized}$ | 5.29253e-7 |

**Table S2.** Ionization cross sections for the stimulated emission field for Mn.

The wave equations for the stimulated emission field are derived within the rotating wave, retarded time($\tau$), and paraxial approximation of Maxwell's equations and are as follows :

$$\left[\frac{\partial}{\partial z} - \frac{i}{2k_0}\nabla_\perp^2 + \mu_s(\mathbf{r},\tau)\right]\begin{pmatrix}\Omega_{s,\text{det}}^+(\mathbf{r},\tau) \\ \Omega_{s,\text{noise}}^+(\mathbf{r},\tau)\end{pmatrix} = i\frac{3}{8\pi}\lambda_0^2\Gamma_{rad}\begin{pmatrix}n(\mathbf{r})\sum_{l,u}T_{lus}\rho_{ul}(\mathbf{r},\tau) \\ f_s(\mathbf{r},\tau)\end{pmatrix} \quad (5)$$

$$\left[\frac{\partial}{\partial z} + \frac{i}{2k_0}\nabla_\perp^2 + \mu_s(\mathbf{r},\tau)\right]\begin{pmatrix}\Omega_{s,\text{det}}^-(\mathbf{r},\tau) \\ \Omega_{s,\text{noise}}^-(\mathbf{r},\tau)\end{pmatrix} = -i\frac{3}{8\pi}\lambda_0^2\Gamma_{rad}\begin{pmatrix}n(\mathbf{r})\sum_{l,u}\rho_{lu}(\mathbf{r},\tau)T_{uls} \\ g_s(\mathbf{r},\tau)\end{pmatrix} \quad (6)$$

$\Omega_s^\pm(\mathbf{r},\tau)$ represents the emitted electric field in units of the rabi frequency with polarization index $s$.

$$\Omega_s^\pm(\mathbf{r},\tau) = \frac{d_0}{\hbar}\mathcal{E}_s\left(\mathbf{r},\tau+\frac{z}{c}\right)e^{\pm i\omega_0\tau}\mathbf{e}_s \quad (7)$$

$\Omega_s^\pm(r,\tau)$ is split into $\Omega_s^\pm(r,\tau) = \Omega_{s,\text{det}}^\pm(r,\tau) + \Omega_{s,\text{noise}}^\pm(r,\tau)$ which represent the deterministic contributions from stimulated emission and stochastic contributions from spontaneous emission.

$\mu_s(r,\tau)$ is the absorption coefficient and $\rho_{pq}(r,\tau)$ is the continuous variable representation of the local atomic density matrices.

$f_s(r,\tau)$ and $g_s(r,\tau)$ are the stochastic source terms for spontaneous emission and are statistically independent of each other.

The atomic density matrices are evolved through the Bloch equations :

$$\frac{\partial \rho_{pq}(r,\tau)}{\partial \tau} = \frac{\partial \rho_{pq}(r,\tau)}{\partial \tau}\bigg|_{\text{inchorent}} + \frac{\partial \rho_{pq}(r,\tau)}{\partial \tau}\bigg|_{\text{unitary}} + \frac{\partial \rho_{pq}(r,\tau)}{\partial \tau}\bigg|_{\text{noise}} \quad (8)$$

Which has incoherent components corresponding to field ionization(that depend on the cross sections summarized in table S1 and S2), radiative transitions and non-radiative decay, unitary components corresponding to evolution of the atomic density matrices from dipole interaction with the emitted field, and noise components for spontaneous emission.

For further details and numerical implementation of the equations on a 3D grid we refer the reader to supplemental reference[8]. Specific details of the 3D grid parameters used in this work are discussed below.

For all simulations a 900x900nm grid with 64x64 grid points was used for the transverse direction. For the single spike simulations, a retarded time window of 5fs was used with a time step of .01 fs. For the SASE simulations the number of time grid points was set for each simulation retarded time window such that time step would be less than or equal to 0.02fs. The method for SASE pulse generation we used is outlined in ref[10] utilizing a typical coherence time of an XFEL[11]. For the jet density simulations, a $z$ integration step size of 10$\mu$m was used, for the polycrystalline density 5$\mu$m was used and the foil density 0.5 $\mu$m.

*Computation of Angular Spectral Properties of Emitted Radiation :*

We compute the angular spectral distribution of the emitted radiation through the following Fourier transform of the retarded time dependent fields :

$$I_s(\omega,\theta_x,\theta_y,z) \propto \Omega_s^+(\omega,\theta_x,\theta_y,z)\Omega_s^-(\omega,\theta_x,\theta_y,z) \quad (9)$$

$$\Omega_s^\pm(\omega,\theta_x,\theta_y,z) = \int \frac{dxdyd\tau}{(2\pi)^3} \Omega_s^\pm(\tau,x,y,z) \exp(\pm ik_0(x\theta_x + y\theta_y) \mp i\omega\tau) \quad (10)$$

$I_s(\omega,\theta_x,\theta_y,z)$ represents the intensity of the field incident with polarization $s$ on the analyzer crystal after propagating through a sample of thickness $z$. For all computed spectra, we summed over the polarization index $s$. We use the notation $\theta_{x/y}$ to note that the dispersive axis of the analyzer can be orientated vertically or horizontally with respect to the sample. As we were considering the effects of amplified spontaneous emission and simulations are started from noise the distinctions between the *x* and *y* axes are arbitrary. For the remainder of the discussion, we will assert the *x* axis as the dispersive axis $\theta_x = \theta$ and the *y* axis as non-dispersive axis $\theta_y = \phi$

*Filamentation of other Samples at LCLS and SACLA without Spectral Broadening :*

We also observed filamentation in our $MnCl_2$ liquid jet and Mn Foil samples utilized at LCLS, which is plotted in figure S3. We have observed such shots at SACLA in the low gain regime and are also illustrated in Figure S3.

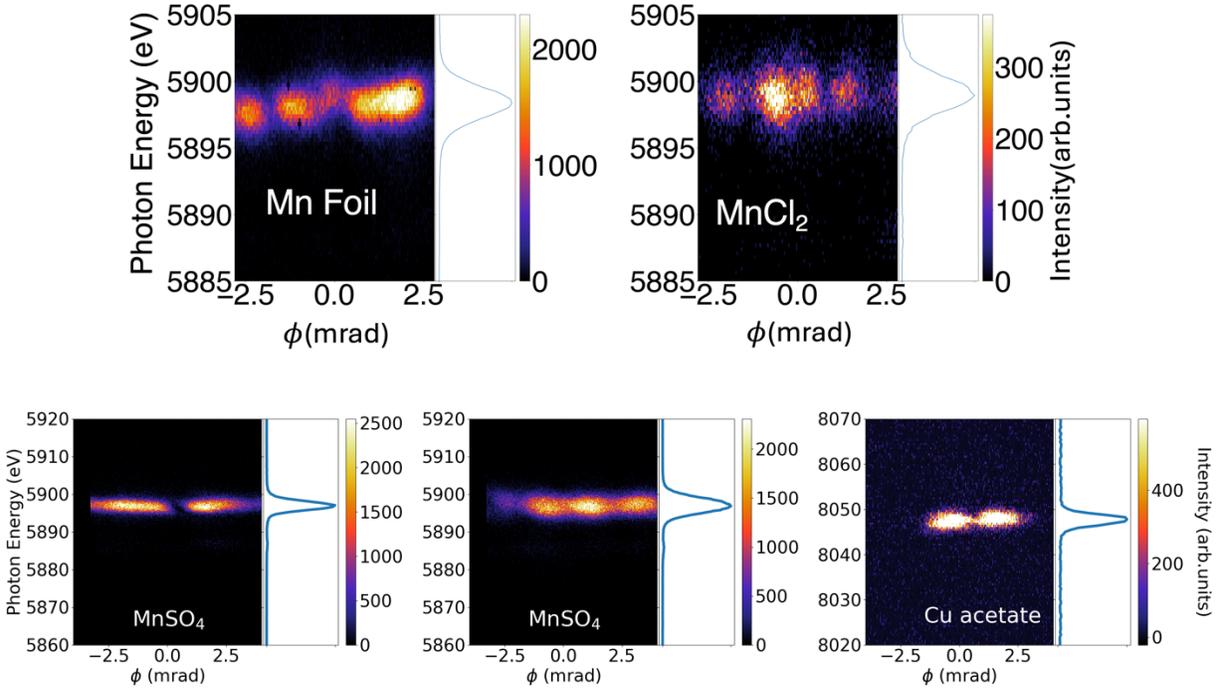

**Figure S3**. Examples of filamentation taken at LCLS and SACLA for different samples.

*Effect of the Analyzer Crystal.*

When the X-ray beam is incident on the Si analyzer crystal each Fourier component of the field is reflected with a coefficient determined by deviation of the field from Bragg condition. The reflected field can be expressed as[2] :

$$\mathcal{E}_R(E,\theta,\phi) = \mathcal{E}(E,\theta,\phi) R\left(\theta + \frac{E}{E_B}\tan(\theta_B)\right) \qquad (11)$$

With $\theta$ and $E$ being the deviations from the central Bragg angle $\theta_B$ and X-ray energy $E_B$ for which the Bragg conditions are satisfied. For a Si 220 analyzer crystal the reflection coefficient has the form

$$R(x) = y - sign(Re(y))\sqrt{1-y^2} \tag{12}$$

$$y(x) = \frac{x \sin(\theta_B)}{\sqrt{\chi_G \chi_{\bar{G}}}} + \chi_0 \tag{13}$$

$\chi_0$ is the uniform component of the X-ray susceptibility, $\chi_G (\chi_{\bar{G}})$ are the periodic components of X-ray susceptibility with reciprocal lattice vectors $G(-G)$.

We plot the magnitude of the response function in Fig. S4

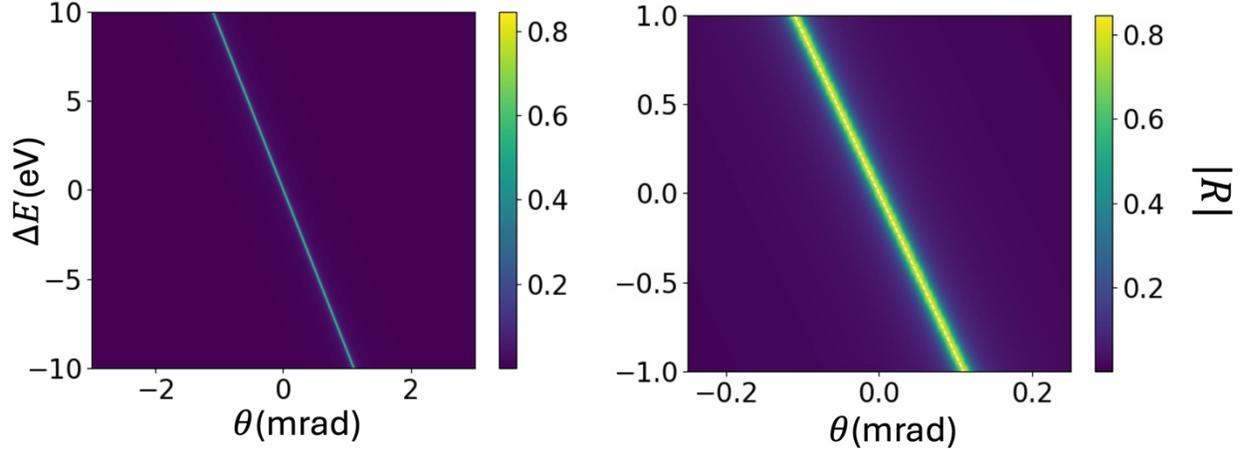

**Figure S4.** Crystal Analyzer Response Function. Dashed line on the RHS represents the DuMond line (eqn. 15)

The detected intensity $I_d(\theta, \phi)$ after reflection by the analyzer crystal can be written as a intergal of the far field intensity $I(E, \theta, \phi)$ with the analyzer response function :

$$I_d(\theta, \phi) = \int dE I(E, \theta, \phi) |R(E, \theta)|^2 \tag{14}$$

The conversion of the detected dispersive axis $\theta$ to photon energy $E_d(\theta)$ (relative to the Bragg energy $E_B$) is done using the Bragg condition for a perfectly reflective crystal:

$$E_d(\theta) = \frac{-\theta E_B}{\tan \theta_B} \tag{15}$$

This condition is commonly referred to as the DuMond line and is plotted as the white dashed line in the right-hand side of Fig. S4.

Due to small spectral and angular width of the response function for the analyzer crystal, for our purposes it can be well approximated as a delta-function along the DuMond line. This implies the detected in intensity can be written as

$$I_d(E_d(\theta), \phi) \sim \int dE \ I(E, \theta, \phi) \delta(E - E_d(\theta)) \tag{16}$$

In addition, for the case where the spectral width is small it implies a very small integration window of over the dispersive axis. For example, if the spectral width is ~3eV the integration window over the dispersive axis is only ~ 0.3mrad for a Si 220 crystal aligned at the Mn $K_{\alpha_1}$ photon energy. As the stimulated emission field varies slowly compared to this integration window, the detected in intensity can be well approximated as a slice of the 3D intensity profile at the dispersive angle corresponding to the $K_{\alpha_1}$ photon energy ($\theta = 0$ for the crystal aligned at this energy) :

$$I_d(E_d, \phi) \sim I(E, \theta = 0, \phi) \tag{17}$$

In Fig. S5 we plot the calculated detected intensity for the filamentation simulation for Mn discussed in the main text using the slice approximation (eqn. S17), the delta function approximation (eqn. S16), and full integral with analyzer response function (eqn. S14).

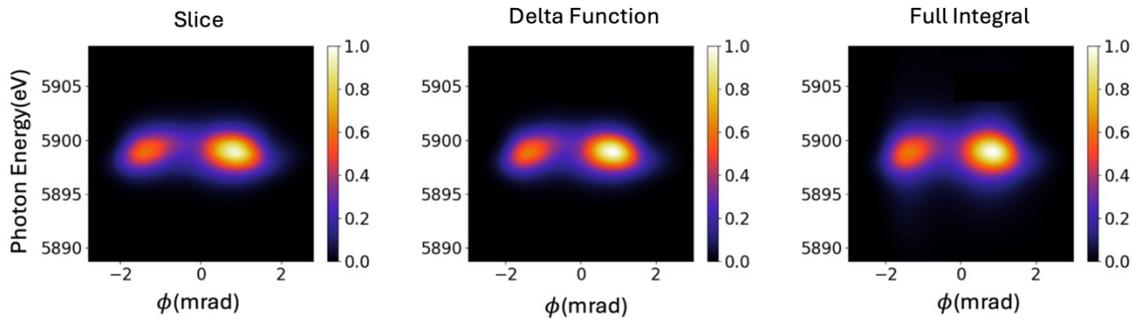

**Figure S5.** Calculation of detected 2D spectral angular profiles using slice (eqn.17 ), delta function (eqn. S16), and full integral (eqn. S14).

The results illustrate the equivalence of the methods for the filamentation spectra with low spectral broadening. For high spectral broadening the slice approximation will breakdown. Throughout the main text and remainder of the supplemental information the delta function approximation was used, with exception of computing the autocorrelation properties of the angular distribution of the filamentation spectra averaged over many stochastic instances, which is discussed later in the supplemental information. In this case the slice approximation was used for computational efficiency.

*Shot to Shot Variations within Filamentation spectra.*

For the vast majority of our LCLS data, we only detect a single amplified mode in the 2D detector. Figure S6 illustrates the distribution of detected hot spots for our NaMnO$_4$ solution data. For 79% of the data, only a single hotspot was detected. As described in the main text, lack hotspot detection does not indicate a lack of filamentation, as we are only sensitive to hotspots along the non-dispersive direction. The hotspot detection was done by passing the raw data through a low pass filter and then applying a peak finder.

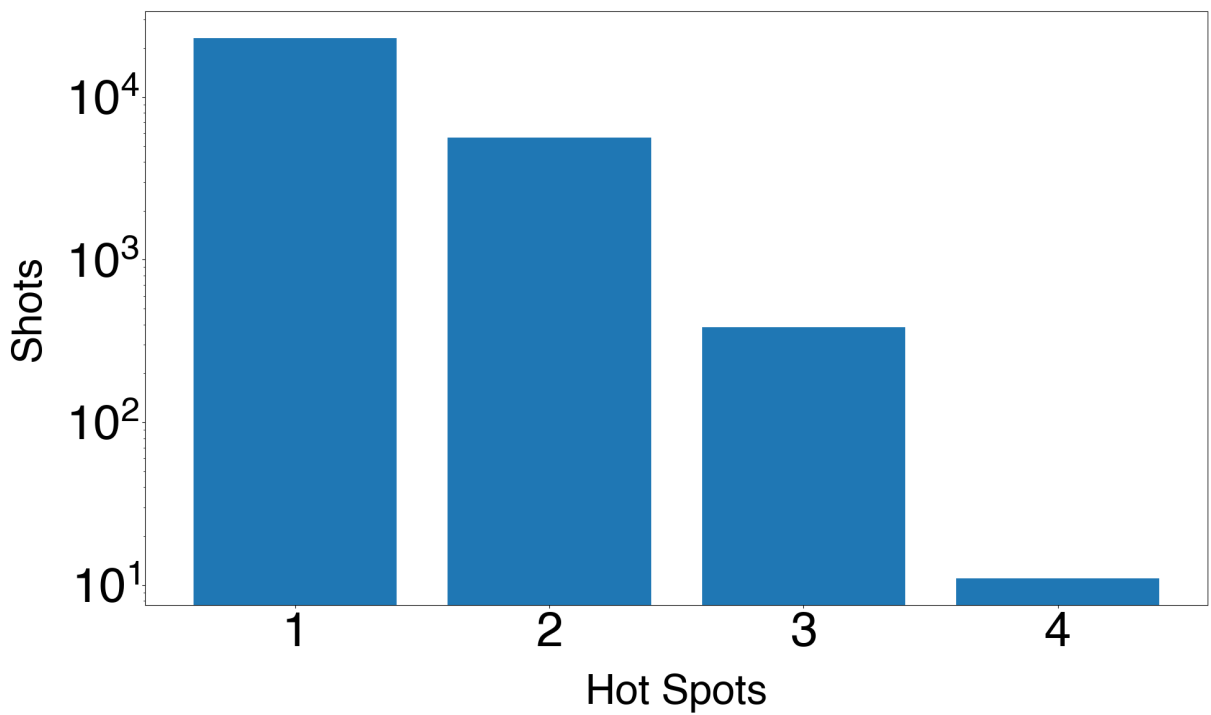

**Figure S6.** Distribution of number of detected hot spots in stimulated emission of NaMnO$_4$ Liquid jets.

Our Maxwell Bloch simulations illustrate that for a given pump pulse there can be large shot to shot variation in the angular distribution of the emitted light, and this is what is observed in experiment. Fig. S7a illustrates the 2D spectral profiles for 9 different stochastic instances of the same single spike pump (~10 uJ FWHM=235 attoseconds) for density equivalent of 2 molar manganese with oxygen and sodium absorbers to intimate a NaMnO$_4$ 2molar solution. Figure S7b illustrates the same for a 20 uJ single spike pump. 9 samples measured for NaMnO$_4$ at LCLS is plotted in Fig. S7c.

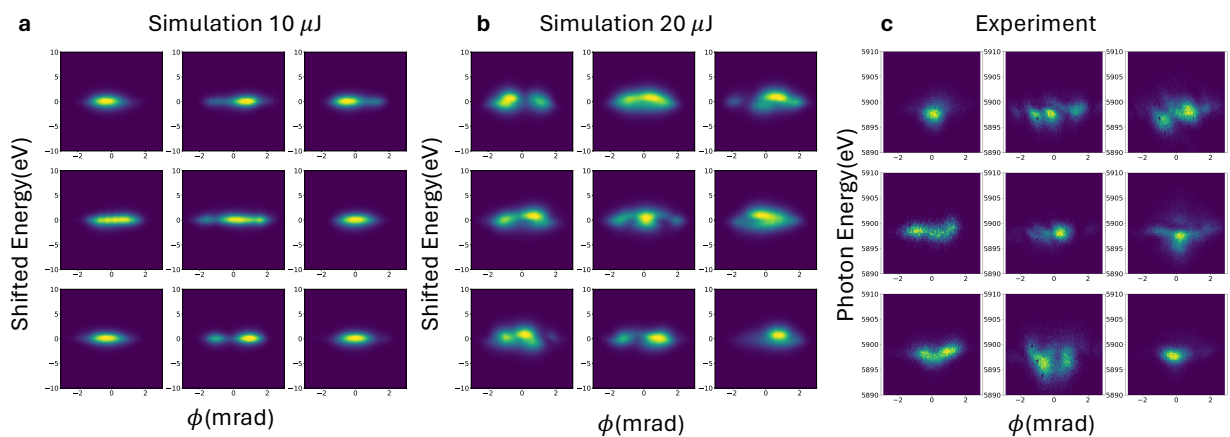

**Figure S7.** Shot to Shot Spatial Variations. (a) Spatial variations for 9 random simulation instances with different seeds for the spontaneous emission process with a 10 $\mu J$ spike pump. Density for simulation was that of 2 molar NaMnO$_4$ (b) Same as (a) but using 20 $\mu J$ spike. (c) 9 Single shot spectra measured for 2 molar NaMnO$_4$ taken at LCLS.

There is clear large variation in spatial distributions of angular hotspots shot to shot for the simulated spectra as well as the experimental spectra. In contrast, for the simulations the spectral properties are reasonably consistent shot to shot for a given pump power. In accordance the temporal profile of the simulated radiation is consistent from shot to shot which is plotted in Fig. S8, with each pulse having a FWHM of about ~1fs for the $10\mu J$ pump and ~0.5 fs for the 20uJ. We note a higher variation for the weaker1 0 $\mu J$ shots. This is consistent with the fact that as stimulated emission approaches saturation the noise terms in our stochastic differential equation approach corresponding to effects of spontaneous emission become negligible compared to the deterministic terms corresponding to the effects of stimulated emission.

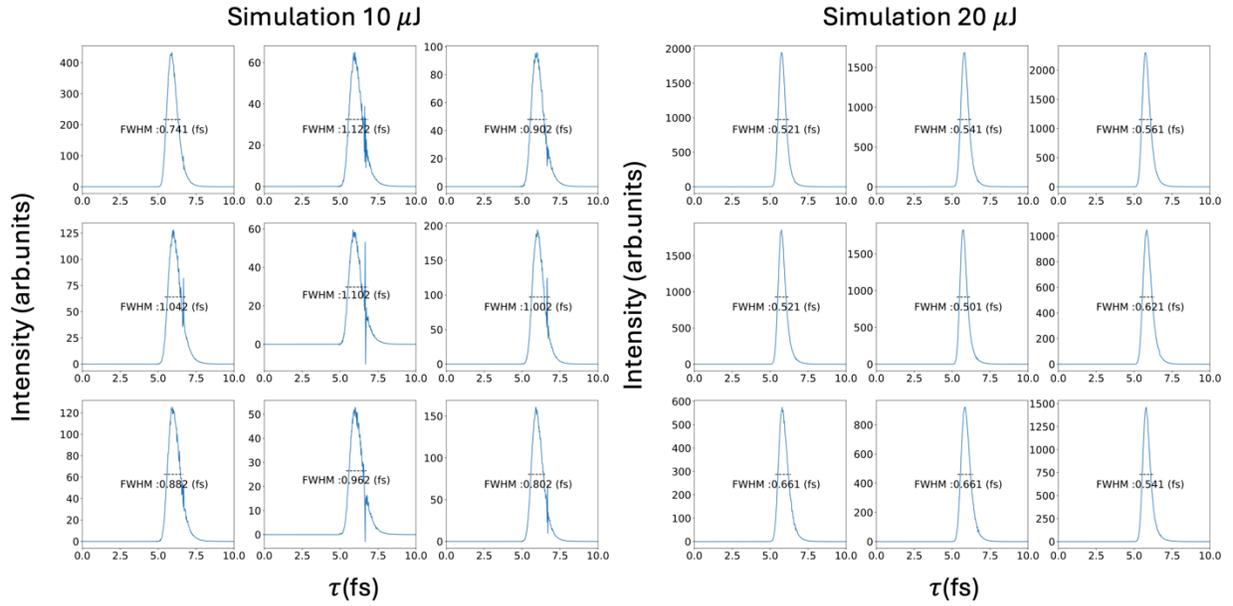

**Figure S8.** Temporal variation shot to shot for stimulated emission spectra.

To compare the ensemble of our simulated angular distribution function to those measured we computed the following distribution and auto-correlation functions :

$$<I_a(\phi')> = <\frac{\int d\phi I(\phi) * I(\phi - \phi')}{\int d\phi I(\phi) * I(\phi)}> \tag{18}$$

$$<I_b(\phi')> = <\frac{I(\phi')}{\max I(\phi')}> \tag{19}$$

Where

$$I(\phi) = \int dE\, I(E, \phi) \tag{20}$$

and <> denotes ensemble averaging. We plot measured and simulated distributions (10 $\mu J$ spike) for 2 Mol NaMnO4 in Fig. S9. While simulations for a single spike pump overestimates the experiment distributions by ~10%, this is reasonable as the within a given run there is a high variation in the effective peak power experienced by the jet due to peak power fluctuations of the SASE pump. It is difficult to improve on this as the proper statistics the SASE pump profile and the precise focusing conditions must be known.

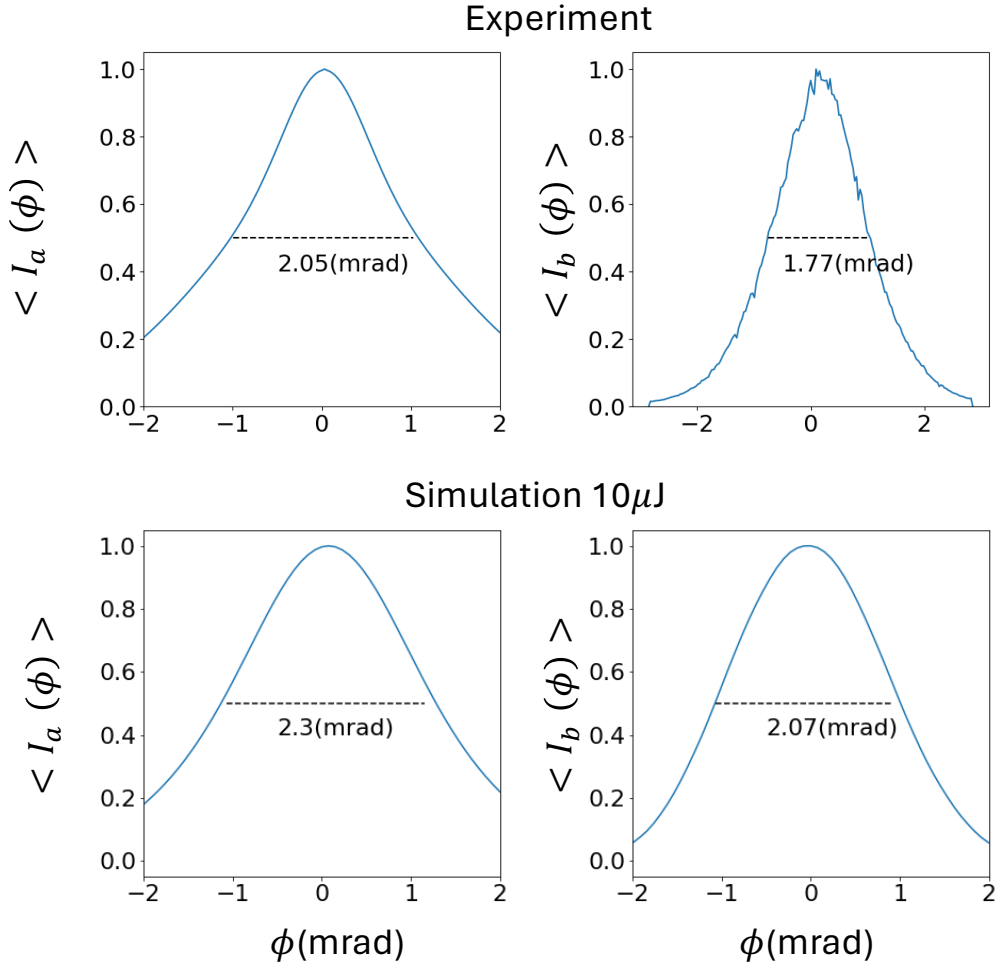

**Figure S9.** Comparison of distribution functions computed with experimental and simulated data.

*Role of the pump pulse profile*

Within our 3D Maxwell Bloch framework, we explored the role of different pump profiles on the emitted $K\alpha_1$ emission. We first considered the role of long (~30fs) and high energy pump pulses (~3mJ) provided at LCLS. Figure S10a illustrates the 2D spectral profile generated by applying a ~30 fs, 3mJ pulse with a noisy SASE temporal behavior. The time profile of the stimulated emission is plotted on top of the pump time profile in Fig, S10b. A zoomed image at the pulse rise is plotted in Fig. S10c. Figures S10b and S10c illustrate the weaker SASE spikes on the pulse rise are strong enough to initiate the stimulated emission process, but not strong enough to initiate highly nonlinear effects.

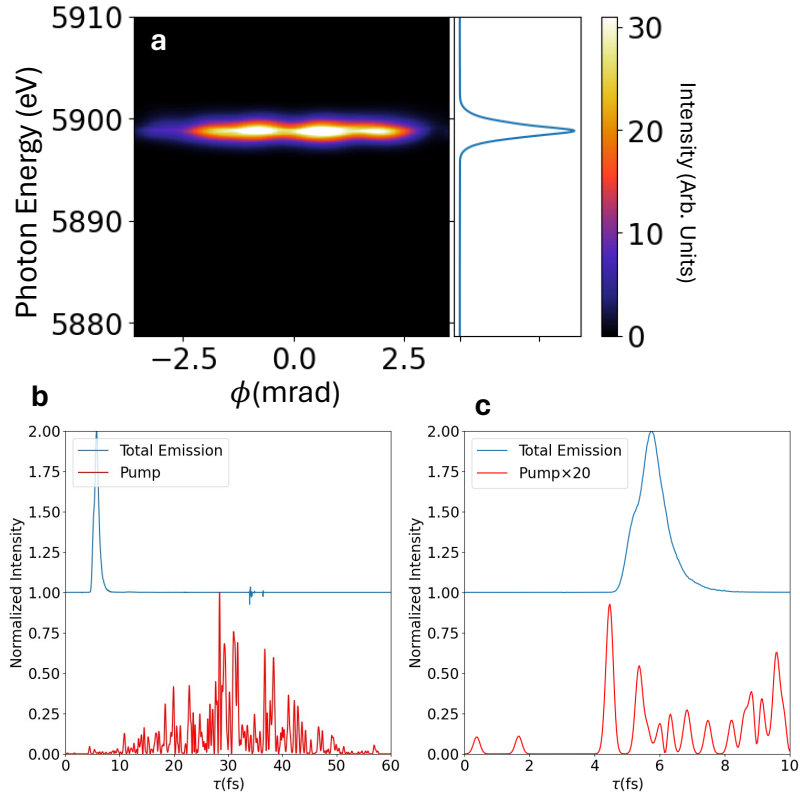

**Figure S10.** Simulation of long pulses. (a) 2D spectra generated by simulation using 30 fs 3mJ SASE pump pulse. (b) and (c) overlay plots of the SASE pump (red) and the stimulated emission (blue). Only weak spikes at the front of the SASE pump pulse initiate the stimulated emission process.

In general, we find the emitted profile is extremely sensitive to the strength of spikes that initiate the stimulated emission process as discussed in the main text.

*Mollow Triplets*

We have also observed Mollow Triplets, which consists of a single peak at the spontaneous emission line and two sidebands that are separated from the main peak by the Rabi frequency. We plot 4 examples of observed Mollow Triplets with Rabi Frequencies greater than 5 eV in Fig. S11.

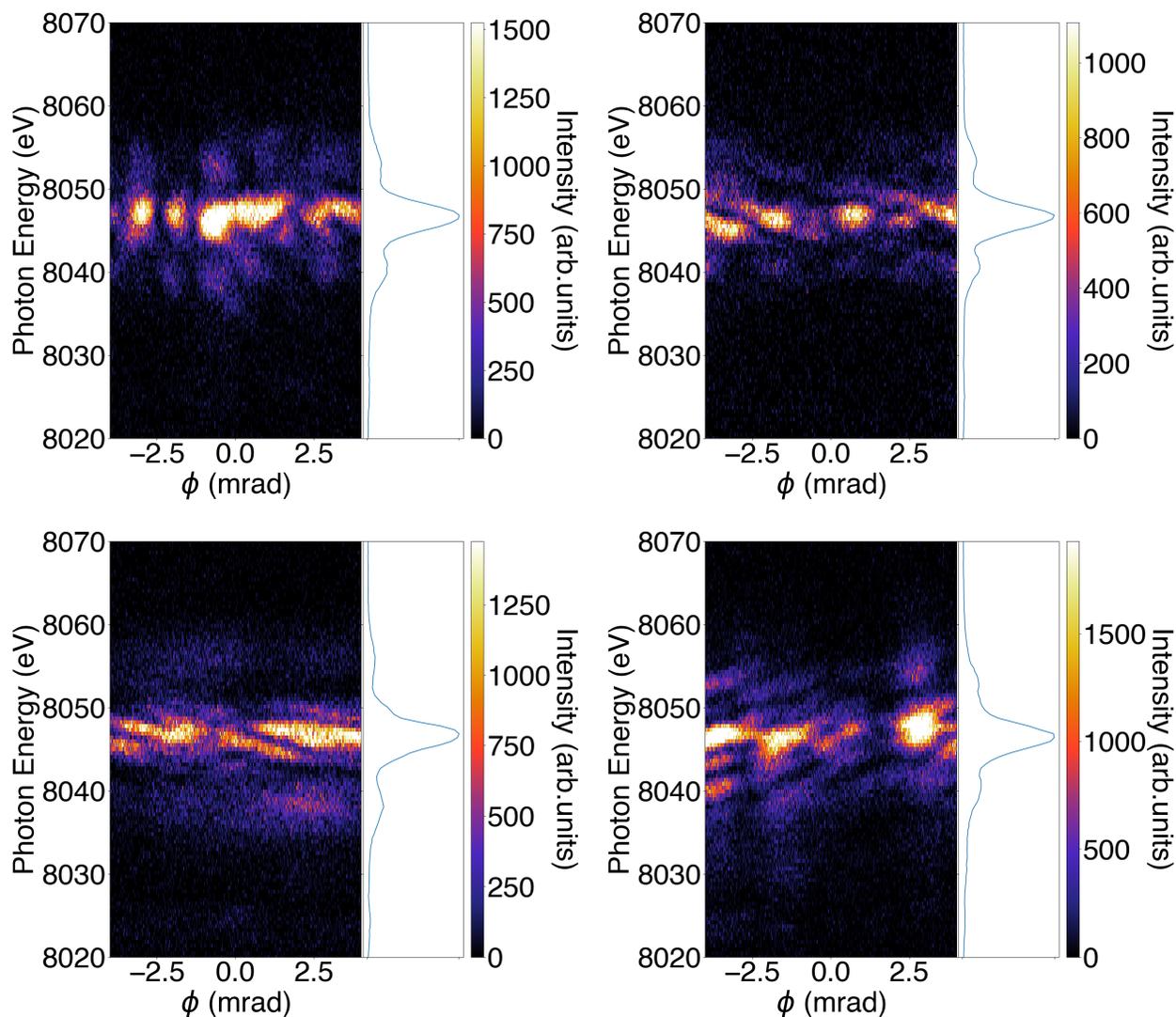

**Figure S11.** Examples of spectra at SACLA resembling Mollow triplets.

The Mollow Triplet will form when there is a continuous population of the excited 1s core hole state from the neutral ground state by the pump pulse in conjunction with Rabi cycling between the between the 1s (*u*) and 2p(*l*) core hole states. Such dynamics can be driven by a SASE pump pulse when multiple spikes continuously depopulate the ground state. A simulated Mollow Triplet spectra is illustrated in Fig. S12a, and the corresponding emission pulse and driving pump pulse in Fig. S12b. Figure S12c illustrates the population dynamics illustrating clear negative population inversion between the 1s (*u*) and 2p(*l*) core hole states, indicating Rabi cycling. Figure S12d shows that during this negative population inversion the ground state is still being depopulated by the pump pulse.

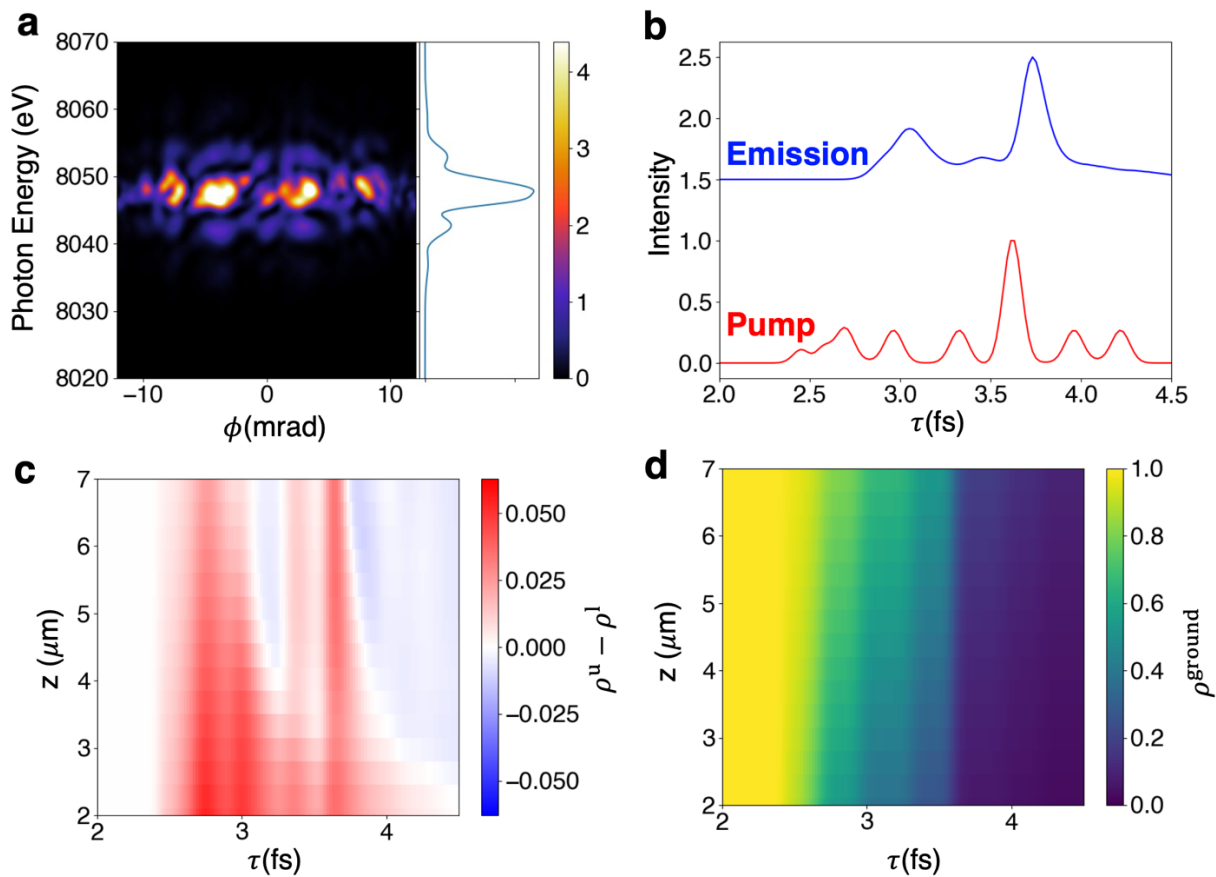

**Figure S12.** Simulation of Mollow Triplet formation. (a) Simulation spectra and (b) corresponding emission profile. Dynamics are driven by multiple spikes in the SASE pump pulse. (c) Population dynamics showing negative population inversion between the upper ($1s_{1/2}$) and lower ($2p_{3/2}$) core hole states indicative of Rabi cycling. While Rabi cycling occurs, the ground state is still be depopulated as shown in (d). This results in the observed Mollow Triplet.

*Multiplets, broadening and Rabi cycling:*

In spontaneous X-ray K$\alpha_1$ emission, spectral broadening is determined by multiplet splitting of the final states and the core hole lifetime[12]. Throughout most of the main text we focus on NaMnO$_4$, as NaMnO$_4$ adopts a formal Mn$^{7+}$ (Mn(VII)) oxidation state leading to a single transition line for the K$\alpha_1$ line. In contrast MnCl$_2$ with Mn$^{2+}$ (Mn(II)) formal oxidation state has several transitions due to atomic multiplet splitting. Figure S13 illustrates this with atomic multiplet calculations of the K$\alpha$ transition for Mn$^{7+}$ and Mn$^{2+}$ ions within O$_h$ symmetry. Calculations were performed using the Green's function implementation outlined in the Quanty software.[13] A small width (0.1eV) for the life-time broadening in the Green's function approach was utilized to illustrate the individual transition lines.

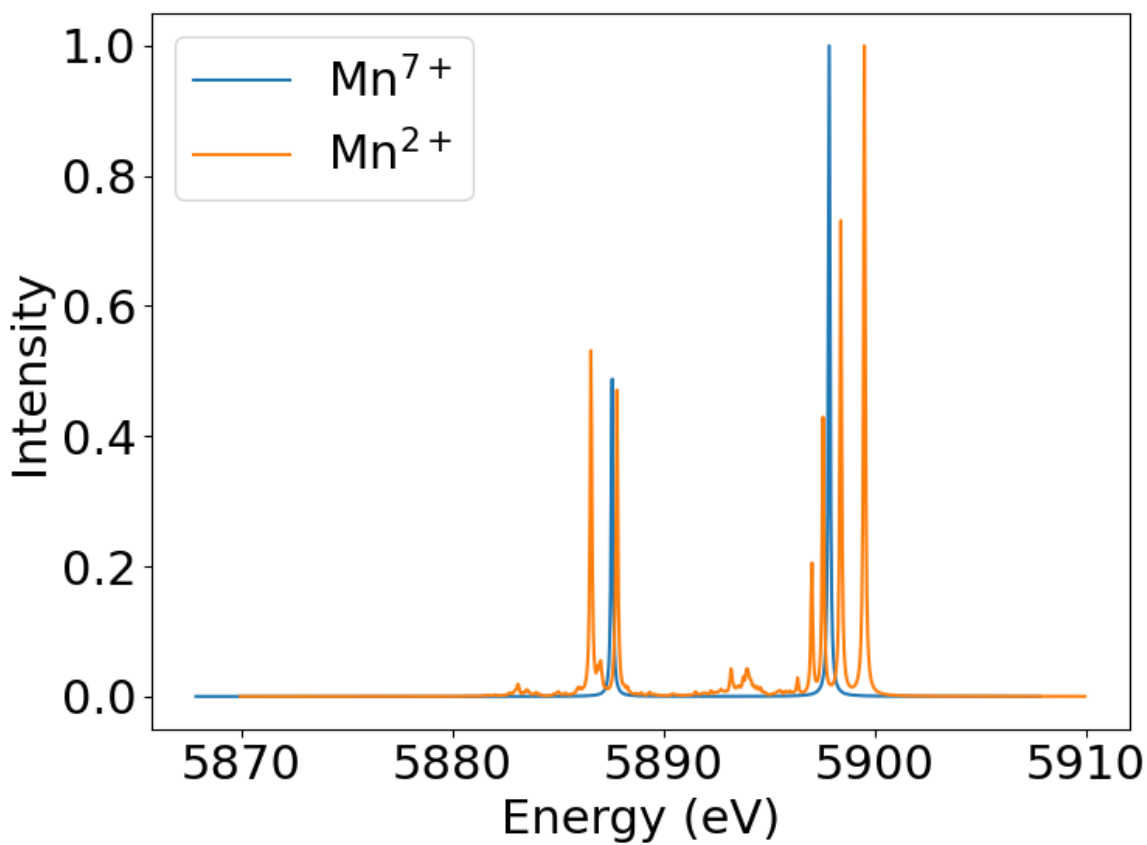

**Figure S13.** Calculated emission spectra for $Mn^{2+}$ and $Mn^{7+}$ ions. $NaMnO_4$ adopts a formal $Mn^{7+}$ oxidation state leading to a single transition line for the $K\alpha_1$ line.

We note previous calculations[1] included ligand-to-metal charge transfer (LMCT) terms to describe Mn-O tetrahedra in $NaMnO_4$, but this only led to a small splitting and shift of the $K\alpha_2$ line which is not crucial to following analysis.

In stimulated emission studies, a *decrease* in spectral width compared to spontaneous emission has been seen in $MnCl_2$[1]. This can happen in the exponential regime of amplified spontaneous emission even when using long high-intensity XFEL pulses. This is due to the spectral $K\alpha_1$ width of $MnCl_2$ being dominated by the multiplet splitting as illustrated in Fig S13. In the exponential regime of stimulated emission, the strongest emission channel, i.e., the transition to a particular final state with the highest oscillator strength in the spontaneous spectrum, is most likely to be amplified first. This can result in a narrowing of the stimulated XES spectrum. Transient amplification schemes, i.e., amplifiers with a rapid gain decay are known to manifest the buildup of transform limited pulses. It has been numerically demonstrated in previous studies that the amplification starts with gain narrowing both in the temporal and spectral domain in the first 1–2 gain lengths. After that, transform limited pulses are created that amplify with almost constant pulse profile, until saturation sets in.

Thus, spectral changes and broadening in Mn and Cu K$\alpha_1$ spectra are dominated by changes to stimulated emission through Rabi-cycling or self-phase modulation versus enhancement of different multiplet lines.

For our simulations a time bandwidth product of ~1.75 eV fs is found in the exponential regime, in response to a single spike pump. For NaMnO4 we observed spectral widths on the order of 1-2.5 eV(see main text figure 3) in the exponential amplification regime. For Cu sulfate we see ~2-3.5eV widths in the exponential regime as plotted in figure S14. These widths indicate sub-fs pulses even in the exponential regime for both Cu and Mn samples.

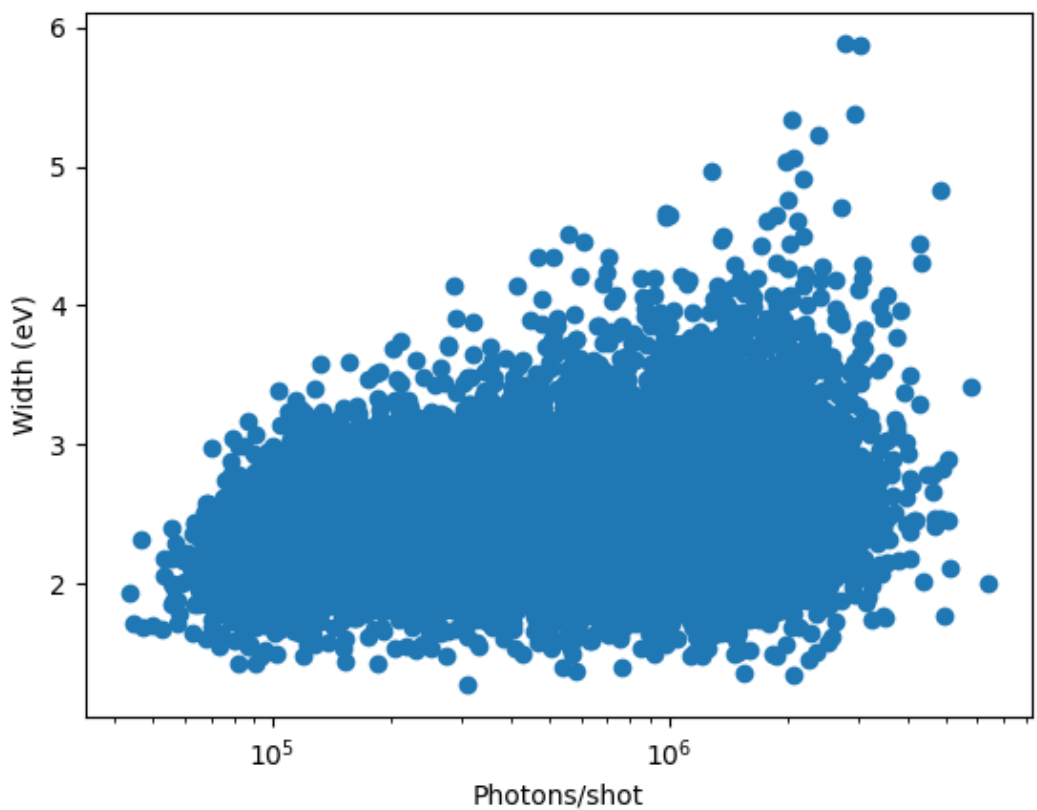

**Figure S14.** Stimulated Emission width versus photon count for Cu Sulfate in low lasing regime.

We note that ionization of the valence *d* electrons can introduce spectral broadening through introducing new resonances corresponding to the different ionization states. However, these effects are small. This has been studied in Fe using high intensity XFEL pulses whereby increasing the pulse length and/or intensity resulted in complete loss of valence to core intensity indicative of complete ionization of valence electrons[14]. In this case only small ~1-2 ev shifts and less than 1eV increase in the spectra width was seen in the K alpha spectra. Across all possible ionization states of Mn in different chemical environments (Mn(VII) corresponding to d0 to high-spin Mn(II) corresponding to d5), the largest observed shifts are less than 2 eV and for Cu, the shifts are much smaller. Ionization of deeper core levels leads to changes in the spectral satellite lines (e.g. KL alpha) that are well outside our observed window (>50 to several 100 eV).

In addition to broadening, we also find a red shift in the average spectrum when averaging over the most intense the shots as illustrated in Figure S15. This red shift is consistent with the predicted self-phase modulation of stimulated emission upon entering the strong lasing regime from 3D maxwell Bloch simulations.

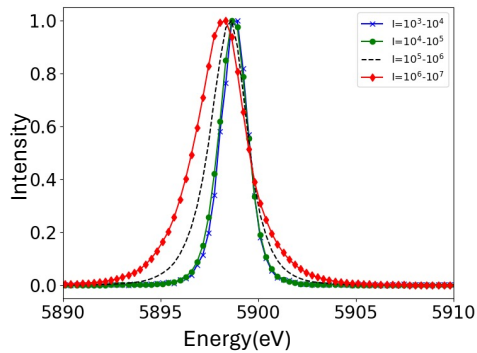

**Figure S15.** Average spectrum in different intensity regimes of the stimulated emission. Intensity is detector units with on 30 detector units corresponding to ~1 photon.

In addition, in the strong lasing regime for NaMnO$_4$ we see appearance of the self induced Autler Townes splitting with more examples given Fig. S16. The Autler Townes splitting is signature of Rabi-cycling and the asymmetry is consistent with the 3D Maxwell-Bloch simulations, as discussed in the main text. Figure S17 re-plots the main-text simulated 2D profile showing spectral splitting as a result of Rabi cycling alongside the corresponding stimulated emission temporal profile. The simulated pulse length is 320 attoseconds FWHM.

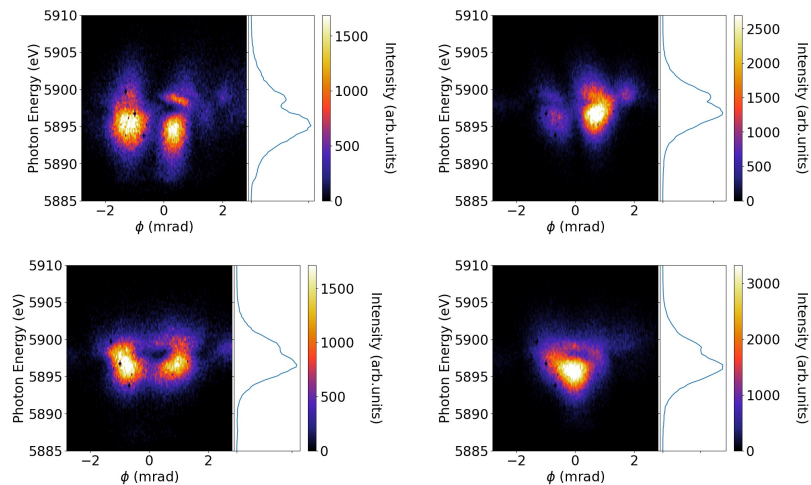

**Figure S16.** Experimental observations of peak splitting of NaMnO₄ spectra upon entering the strong lasing regime indicative of rabi-cycling.

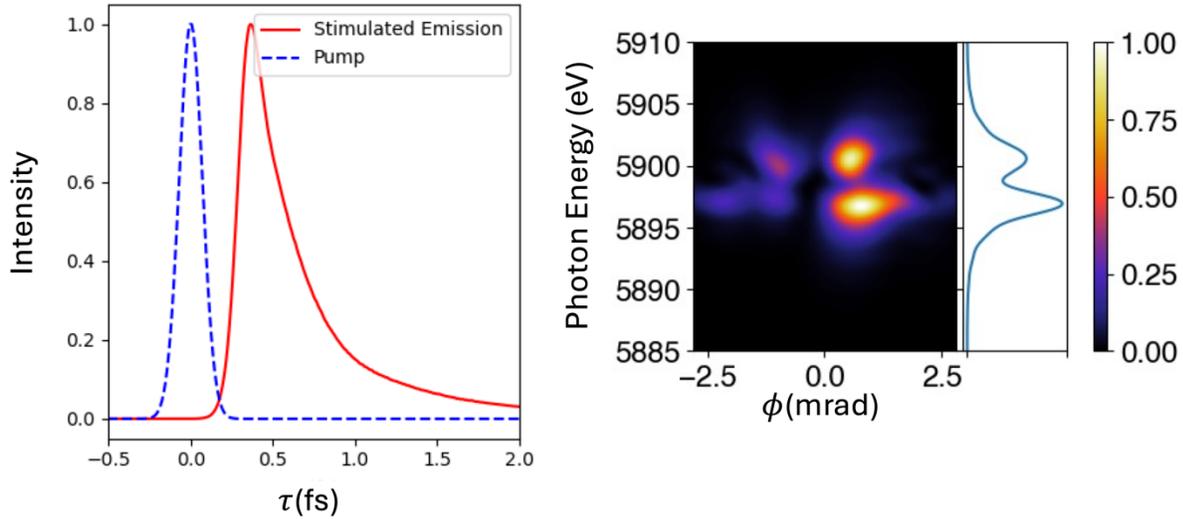

**Figure S17.** Simulation (same as main text figure 3) of self-induced Autler Townes splitting of the stimulated emission. Total pulse length is 320 attoseconds FWHM.

*Sub-fs Autler Townes Splitting*

Figure S18a illustrates an observed Autler-Townes splitting of ~5eV illustrated for a 7$\mu$m copper foil. The simulated response of a 7 $\mu$m Cu foil to a single strong spike showing similar spectral shape is illustrated in Fig. S18b. The corresponding simulated pulse length is 230 attoseconds FWHM, and its temporal profile is plotted in Fig. S18c.

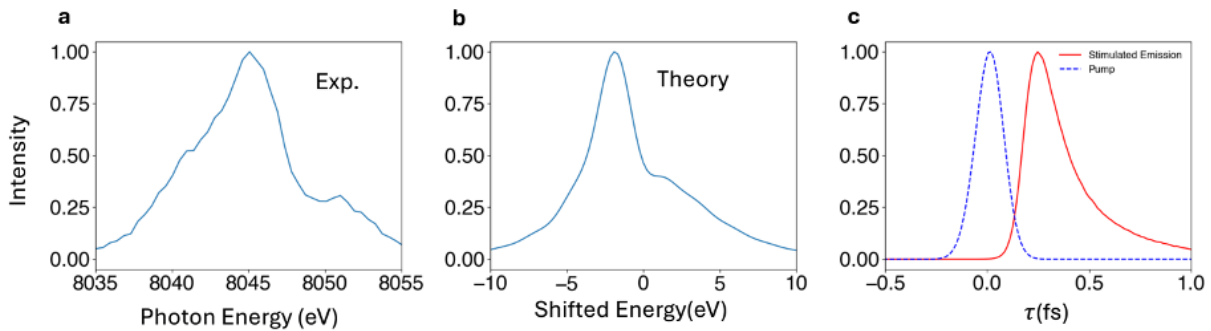

**Figure S18.** Sub-fs Autler Townes splitting. (a) 7 $\mu$m copper foil displaying >5eV Autler Townes splitting in the stimulated emission integrated spectra. (b) Simulation displaying the same spectral shape as (a). (c) Corresponding

*Phase stability of stimulated emission pulses.*

Figures S19a and S19b shows the simulated intensity and phase of the emission pulse at center of the pump focus after 100 μm propagation through a Mn sample in the high gain regime when Rabi cycling occurs. Figures S19a and S19b illustrate that for the main stimulated emission spike the phase is nearly constant within its FWHM indicating it is nearly transform limited. Figure S19c illustrates the transform limit and computed spectra for the full field which red-shifting and additional frequencies appear due to the self-phase modulation of the side bands. In contrast the main emission peak is only slightly blue-shifted compared to the transform limit as illustrated in Figure S19d.

The overall phase stability of the main emission peak is what allows separate stimulated emission events initiated by different SASE spikes in a pump pulse to generate fringe interference patterns in their observed spectral profile. In contrast the multiple side bands resulting from the rabi-oscillations have linearly decreasing phases indicative of their frequency modulation that overall red-shifts the stimulated emission. This frequency modulation results in the spectral inhomogeneities observed in the experiment.

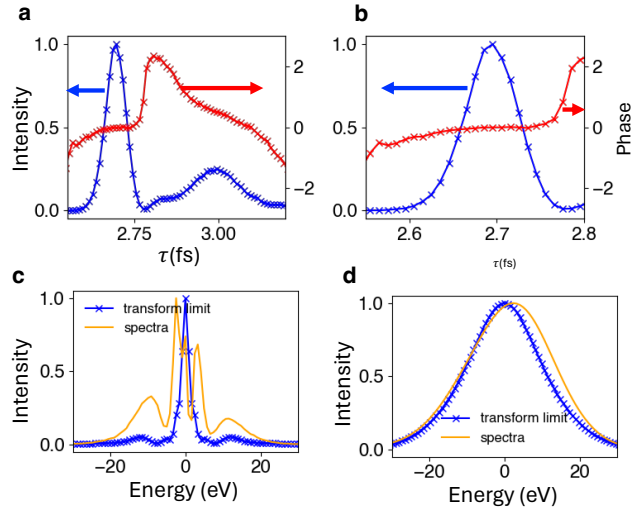

**Figure S19.** Phase Stability of stimulated emission pulses. (a) Intensity and phase for simulated stimulated emission pulse generated by single spike pump propagating through 100 μm thick Mn sample with density 6/nm³. (b) Shows so zoom in on main stimulated emission spike which is phase stable. Comparison of transform limit and computed spectra for the full stimulated emission profile is illustrated in (c) and the main the peak (b).

*Frequency of Strong Lasing Shots and Single Temporal Modes.*
To quantify the frequency of strong lasing shots for copper foils we examine the frequency of shots that can be described well by single gaussian fits ($R^2 > 0.95$), indicative of lasing from a single temporal mode. For Copper 20 μm foils we find only 22% of shots can be described by lasing from a single temporal mode with a width less than twice the spontaneous emission width ($\Gamma_{sp} \sim 2.5 eV$), i.e a width less than 5eV. We have plotted 20 random samples (the first observed 20 shots in a single run) in Fig S20, all displaying broad an inhomogeneous spectra

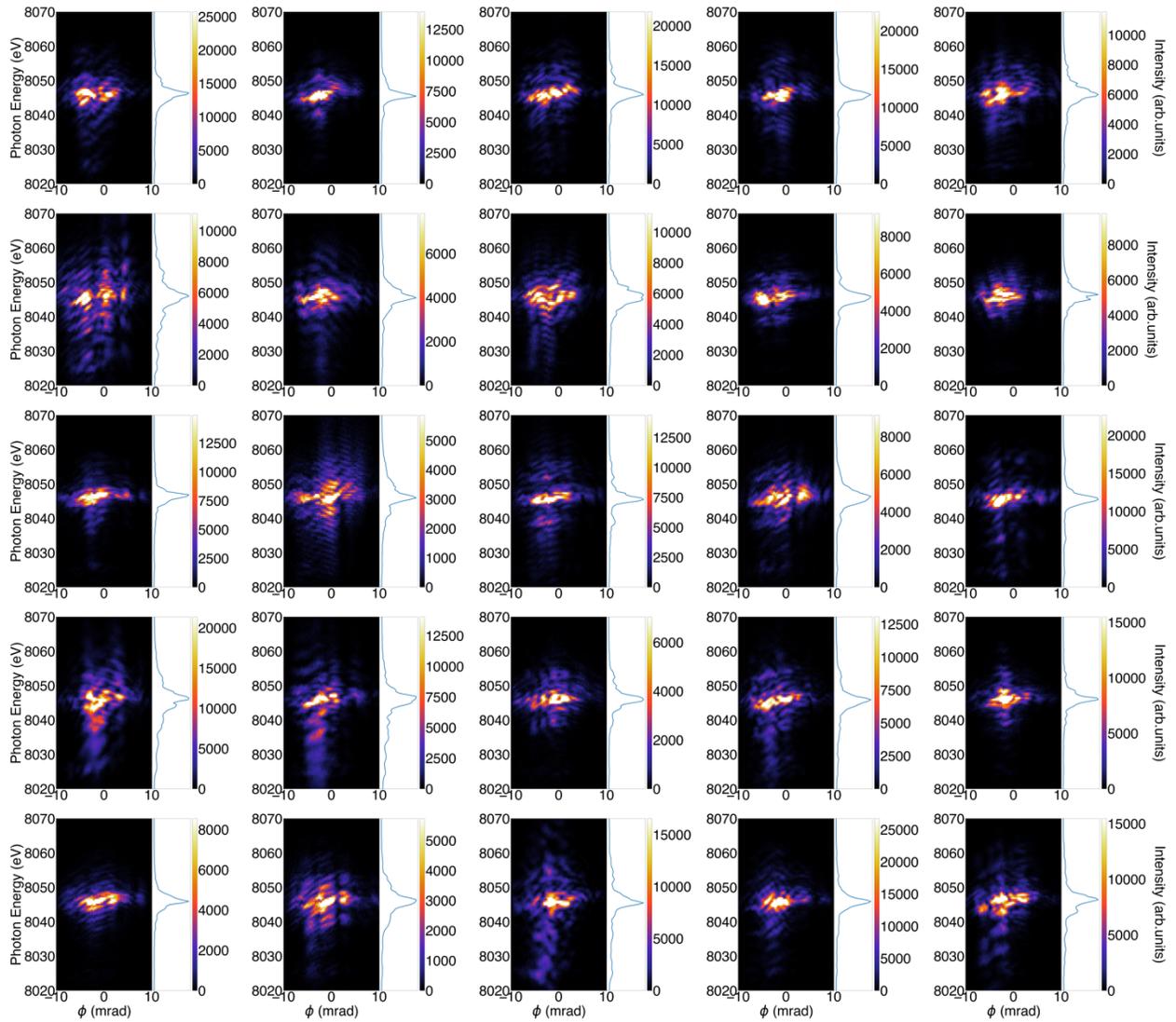

**Fig S20**. 25 random Strong Lasing Shots for Cu 20 $\mu$m Foils.

While the broadest and most inhomogeneous shots occurred with the 20 $\mu$m Foil samples, nearly all contained some amount of fringe contrast indictive of generation of multiple pulses. We find with a 7 $\mu$m Cu foil one can generate single spike pulses ($R^2$ >0.95), with width greater than 5eV (<350 attoseconds based on our simulations for the exponential regime) 15% of the time. Similar results can be achieved through attenuation; however, we found the 7 $\mu$m Cu foil to produce the highest percentage of these types of shots.

*Spectral Fringes*
Well-defined spectral fringes with nearly constant fringe spacings have been previously observed at SACLA. These result from the interference of temporally coherent K$\alpha$ superfluorescencent pulse pairs that were created from population inversion from separate SASE spikes within the pump pulse.[2] We now report observation of coherent fringes with 8.0, 7.1, and 5.4eV spacing indicating sub-fs pulse separation. These shots are illustrated from left to right in supplemental

Fig. S21. For such high contrast fringes to be generated by the interference from coherent pulse pairs their pulse lengths must be much shorter than their corresponding temporal spacing, indicating attosecond pulse pair creation.

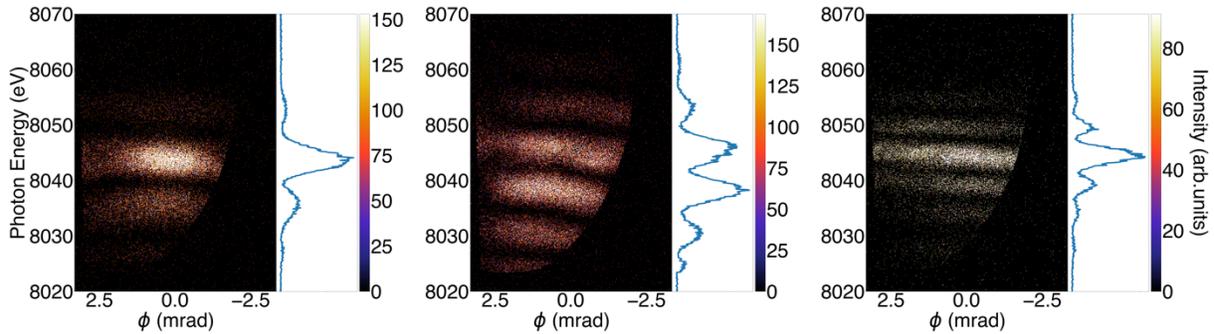

**Fig S21**. Detected deep spectral fringes at SACLA. Data was taken with Si 333 analyzer crystal.